\documentclass{article}
\usepackage{apacite}\newcommand{\url}[1]{}
\usepackage{times}
\usepackage{bm,amsfonts,amssymb,amsmath}
\usepackage{graphicx}
\let\mode=1
\begin{document}

\title{Indisputable facts when implementing \\ spiking neuron networks}

\author{Bruno Cessac$^{1,3}$, H\'el\`ene Paugam-Moisy$^{2}$, Thierry Vi\'eville$^{4}$ \\ 
	{\scriptsize $^{1}$LJAD {\footnotesize \tt \em http://math1.unice.fr} \& }\\
	{\scriptsize $^{2}$ INRIA TAO \& $^{3}$INRIA NeuroMathComp \& $^{4}$INRIA Cortex {\footnotesize \tt \em http://www.inria.fr}}}

\maketitle

\begin{abstract}

``Spikes are the neural code'': this claim is about 15 years old \cite{shadlen-newsome:94,rieke-etal:96}, 
preceded by theoretical studies on the underlying mathematical processes (e.g., \cite{gerstein-mandelbrot:64}), 
and followed by many developments regarding biological modelling or computational paradigms, or both (e.g., \cite{thorpe-delorme-etal:01}).
However the involvement of spikes in neural coding is still an open subject.
Several fundamental aspects of dynamics based on spike-timing have been very recently clarified, both at the neuron level \cite{touboul-brette:08} and the network level \cite{cessac-vieville:08}.
Nevertheless, still a non negligible set of received ideas, as, e.g., the ``incredible power of spikes'' or, e.g., the ``mystery of the [spike based] neural code'' (sic !) are currently encountered in literature. 

In this article, our wish is to demystify some aspects of coding with spike-timing, through a simple review of well-understood technical facts regarding spike coding. The goal is to help better understanding to which extend computing and modelling with spiking neuron networks can be biologically plausible and computationally efficient. 

We intentionally restrict ourselves to a deterministic dynamics, in this review, and we consider that the dynamics of the network is defined by a non-stochastic mapping. This allows us to stay in a rather simple framework and to propose a review with concrete numerical values, results and formula on 
(i) general time constraints, (ii) links between continuous signals and spike trains, (iii) spiking networks parameter adjustments.

When implementing spiking neuron networks, for computational or biological simulation purposes, it is important to take into account the indisputable facts here reviewed.
This precaution could prevent from implementing mechanisms meaningless with regards to obvious time constraints,
or from introducing spikes artificially, when continous calculations would be sufficent and simpler.
It is also pointed out that implementing a spiking neuron network is finally a simple task, unless complex neural codes are considered.
 
\end{abstract}

\noindent {\bf\normalsize Key Words} Spiking neuron network. Neural code. Time constraints. Spike train metrics.

\section{Introduction}

Let us consider, for instance, biological models of cortical maps \cite{koch-segev:98,dayan-abbott:01}, 
in a context where the spiking nature of neurons activity is made explicit \cite{gerstner-kistler:02b}, 
either from a biological point of view or for computer simulation. 
From the detailed Hodgkin-Huxley model \cite{hodgkin-huxley:52}, 
(still considered as the reference but unfortunately intractable when considering neural maps), 
back to the simplest integrated and fire (IF) model, a large family of continuous-time
models have been produced, often compared with respect to their (i) biological plausibility and their (ii) simulation efficiency.

Theoretically, spiking neurons can perform very powerful computations with precise spike timings.
Spiking neurons are at least as computationally powerful as the sigmoidal neurons traditionally used in artificial neuron networks \cite{maass:97,maass-natschlager:97}.
This result has been shown using a spike-response model (see \cite{maass-bishop:03} for a review) and considering piece-wise linear approximations of the membrane potential profiles.
In this context, analog inputs and outputs are encoded by temporal latencies of spike firings.
It has been shown that any feed-forward (multi-layer) or recurrent analog neuronal network (e.g. Hopfield network) 
can be simulated arbitrarily closely by an insignificantly larger network of spiking neurons. The assertion holds even in the presence of noise \cite{maass:97,maass-natschlager:97}. 
Such theoretical results highly motivate the use of spiking neuron networks for modelling and simulation purpose.

\subsubsection*{Biological plausibility of neuron network models.}

 Biological plausibility at the neuron level is understood as the ability to reproduce what is observed at the cell level, often considering in-vitro experiments 
\cite{koch-segev:98}. The point of view is questionable as shown in recent experiments in V1 \cite{fregnac:03,fregnac:04} where it appears that a single-cell observation
highly differs between in-vitro and in-vivo conditions. 
Biological plausibility at the network level is understood as the ability to reproduce what is observed regarding e.g. the cortical map 
activity \cite{carandini-demb-etal:05}. This includes predicting the response not only to specific artificial, but also 
natural stimuli: this means, for V1, taking into account
natural image sequences input shifted by eye movements \cite{baudot-marre-etal:07}, after the retinal and LGN processing 
(see e.g. \cite{simoncelli-olshausen:01} for a discussion about information processing in these structures).

 As far as this contribution is concerned, we consider a weaker notion of biological plausibility: A simulation is biologically plausible if it verifies an explicit
set of constraints observed in biology. More precisely, we are going to review and discuss a few time constraints, shared by all dynamics, 
further called ``general time constraints''. We develop their consequences at the simulation level.
The time constraints are based on biological temporal limits and appear to be very precious quantitative elements, both for estimating the coding capacity of a system and for improving simulations.

\subsubsection*{Simulation efficiency of integrate and fire models.}

Among all the spiking neuron models, the punctual conductance based generalized integrate and fire (gIF) is an adaptive, bi-dimensional, non-linear, integrate-and-fire model with conductance based synaptic interaction
(as e.g. in  \cite{destexhe:97,brette-gerstner:05,rudolph-destexhe:07}).
At the present state of the art, considering gIFs as neuron models
presents several advantages: 

- They provide an effective description of the neuronal activity allowing one to reproduce several important neuronal regimes \cite{izhikevich:04}, 
well matching to biological data, especially in high-conductance states, 
typical of cortical in-vivo activity \cite{destexhe-rudolph-etal:03}. 

- Nevertheless, they consist of a simplification of Hodgkin-Huxley models, which is useful both for mathematical analysis and numerical 
simulations \cite{gerstner-kistler:02b,izhikevich:03}.

In addition, though these models have mainly been considered for studying the dynamics of a single neuron, they are easy to extend to a network structure,
including synaptic plasticity modelling \cite{markram-etal:97,pfister-gerstner:06}.

See, e.g. \cite{rauch-etal:03} for further elements in the context of experimental frameworks and \cite{la-camera-etal:08a,la-camera-etal:08b} for a review.

However, in all the variants of integrate and fire models, it is assumed that an {\em instantaneous} reset of the membrane potential occurs after each spike firing, except for the Spike Response Model of \cite{gerstner-kistler:02b}.
The reset is a formal simplification and has a general spurious effect: Information theory (e.g. Shannon's 
theorem, stating that the sampling period must be less than half the period corresponding to the highest signal frequency) is not applicable with unbounded frequencies. 
From the information theory point of view, it is a temptation to relate this spurious property to the {\em erroneous} fact 
that the neuronal network information is not bounded.
In the biological reality, time synchronization is indeed not instantaneous (action potential time-course, synaptic delays, refractoriness, ...).

\subsubsection*{What is the paper about}

In section~\ref{time-constraints}, we emphasize the fact that, in computational or biological contexts, 
not all time sequences correspond to realistic spike trains since they are constrained by the neural dynamics, while general time constraints are also to be taken into account.
We revisit this apparently obvious point and provide numerical evaluations. One of the constraints we propose is far from being obvious and we discuss that point in detail.

In section~\ref{info-amount}, we make explicit the maximal amount of information present in a ``true'' spike train, 
taking the general time constraints into account. This point of view contradicts what is usually implicitly assumed
about the ``incredible power of computating with spikes'' and, although obvious, it is worth reminding us about the limitation we derive.

In section~\ref{about-dynamics}, we review a recent work which clarifies the kind of dynamics encountered in deterministic integrate and fire neuron networks,
demystifying the notion of ``chaotic spiking network dynamics'' and supplying a rigorous notion of what is called the ``edge of chaos''.

In section~\ref{neural-coding}, we discuss to which extend defining the ``neural code'' contained in spike trains is related to the choice of a metric, 
in the deterministic case, i.e. when the dynamics of the neuron network is defined by a non-stochastic mapping.
The relation with existing neural codes (rate coding, rank coding, phase coding, ...) is discussed.

As a first major consequence, considering convolution metrics in section~\ref{continous-metrics}, we can make explicit, in the linear case, 
the links between spike trains and continuous signals, with concrete methods to build such a link.

As a second major consequence, considering alignment metrics in section~\ref{alignment-metrics}, we can describe methods to explicitly program
spiking neuron network parameters in order to obtain a given input/output relation in the deterministic case again.

\section{General time constraints in spike trains}\label{time-constraints}

The output of a spiking neuron network is a set of events, defined by their occurrence times, up to some precision:
\\ \centerline{${\cal F} = \{ \cdots t_i^n \cdots \}$ \ \ \mbox{with}\ \  $t_i^1 < t_i^2 < \cdots < t_i^n < \cdots,\  \forall i, \ \forall n$}  \\
where $t_i^n$ is the $n$th spike time of the neuron $i$, with related inter-spike intervals $d_i^{n} = t_i^{n} - t_i^{n-1}$. 
 \if 2\mode Such {\em spike train} writes $\rho_i(t) = \sum_{t_i^n \in {\cal F}_i} \delta(t - t_i^n)$, using the Dirac symbol $\delta(.)$.\fi
~See e.g. \cite{dayan-abbott:01,gerstner-kistler:02b,schrauwen:07,paugam-bohte:09} for an introduction to spiking neuron networks.

In computational or biological contexts, not all sequences ${\cal F}$ correspond to spike trains since they are constrained by the neural dynamics.
In computational or biological contexts, the following time constraints must be taken into account:

\begin{itemize}
\item [[C1]] The inter-spike intervals are bounded by a refractory period $r_i$, $d_i^n > r_i$,
\item [[C2]] The spike times are defined up to some absolute precision $\delta t$
\item [[C3]] There is always a minimal delay $dt$ for a pre-synaptic spike to influence a post-synaptic neuron, thus having a causal effect on another spike
\item [[C4]] There is a maximal inter-spike interval $D$ such that $\forall i, \ \forall n$ either $d_i^n < D$ or $t_i^n = +\infty$\\ (i.e. either a neuron fires within a time delay $< D$ or it remains quiescent forever).
\end{itemize}

For biological neurons, orders of magnitude are typically, in milliseconds: 
$$
\scriptsize \begin{tabular}{|c|c|c|c|} \hline $r$ & $\delta t$ & $d t$ & $D$ \\ \hline $1$ & $0.1$ & $10^{-[1, 2]}$ & $10^{[3, 4]}$ \\ \hline \end{tabular}
$$
These numerical evaluations are discussed in the present section.\\

The [C1] constraint is well-known as a limit for the maximal firing rate. See e.g. \cite{koch:99b} for an extended discussion on absolute / relative refractory periods.\\

The [C2] constraint may correspond to more than one definition. 
For instance, probabilistic interpretations often consider an additive perturbation in the dynamic evolution, 
to encounter for the fact that spike times are not precisely defined.
On the other hand, deterministic interpretations may consider precision intervals. Here, we propose a simple deterministic specification:

\begin{center}
\textit{Two spike times are different, e.g., not synchronized, if separated by more than $\delta t$.}

\textit{Two spike times are indistinguishable if they are separated by less than $\delta t$. }
\end{center}
 
Indistinguishable does not mean ``equal'', but that means we can not state if equal or different.
With such a deterministic interpretation, $\delta t$ can be calculated using 1st order approximations.
The [C2] constraint is sometimes ``forgotten'' in models. 
In rank coding schemes for instance \cite{gautrais-thorpe:98} it is claimed that ``all'' spike-time permutations are significant, 
which is not realistic since many of these permutations are indistinguishable, because of the bounded precision, as discussed in e.g. \cite{vieville-crahay:04}. 
Similarly, a few concepts related to ``reservoir computing'' (see e.g. \cite{paugam-moisy-etal:08} and quoted contributions, for a review) 
do not address this issue, although simulations indeed have to take it into account. As a consequence, an unrealistic unbounded time precision is implicitly assumed. 

\begin{quotation} { \subsubsection*{Spike time precision evaluation}

Considering that the spike time of a real neuron is defined by the time $t_i$ when the membrane potential $V(t_i)$ reaches a maximum, 
we obtain around $t_i$, assuming differentiability of $V$:
$$V(t) = V(t_i) + \kappa \, (t - t_i)^2 + o(|t - t_i|^2)$$
with $\kappa = d^2 V/d t^2(t_i)$ and easily derive, as a rule of thumb for the spike-time precision $\delta t$:
$$\delta t \simeq \sqrt{\frac{<\delta V>}{<\kappa>}}$$
where $\delta V$ is the voltage precision and the averages $< >$ are to be taken over a set of measurements. This formula is derived from standard 1st order error analysis.

In order to roughly estimate spike time precision, we have considered a few dozen of spike profiles in several spike trains \cite{carandini-ferster:00,koch:99b} and
graphically estimated the values in a zoom of the provided figures. We have obtained $\delta t \simeq 0.1 ms$, with 
a peak curvature order of magnitude $<\kappa> = 100 mV/ms^2$ as illustrated in Fig.~\ref{koch}, considering
a voltage precision of average $<\delta V> = 10 \mu V$, i.e. at the order of magnitude of the membrane potential noise \cite{koch:99b}. 
Similar numerical values are obtained reading other electro-physiological data \cite{carandini-ferster:00}.
 \begin{figure}[htb] 
   \centerline{\includegraphics[width=0.7\textwidth,height=4cm]{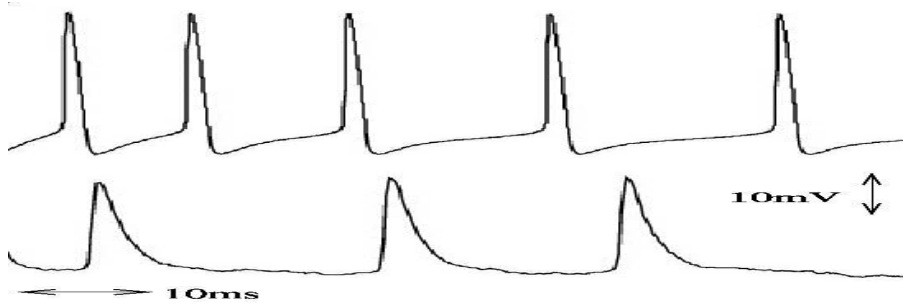}}
   \caption{\label{koch} Two examples of spike profiles in the cat primary visual cortex. The peak curvature order of magnitude are $30-100 mV/ms^2$.}
 \end{figure}

Furthermore, a similar order of magnitude is obtained in literature, considering the numerical precision in inter-neuron synchronization \cite{crook-ermentrout-etal:98}
which is found of about $1 ms$, while \cite{mainen-sejnowski:95} (e.g. in Fig.2B) report submillisecond accuracy in vitro, but no higher than $0.1$ms.

} \end{quotation}

Similarly, [C3] is obvious and has for consequence to avoid every spurious effects\footnote{If a neuron instantaneously fires after receiving a spike, 
this can generate avalanche effects (another neuron instantaneously fires and so on) or 
even temporal paradoxes (another inhibitory  neuron instantaneously fires inhibiting this one, thus not supposed to fire any more).}, 
and induce simplifications both at the modelling and simulation levels
\cite{morrison-mehring-etal:05}.

\begin{quotation} { \subsubsection*{Spike time propagation evaluation}

Delays from one spike to another involve the pre-synaptic axonal delay, the synaptic delay and the post-synaptic dendritic delay. The smaller observed delays \cite{koch:99b,burnod:93} seem to be at least of $0.5 ms$, with values up to $40-50 ms$ for inter cortical maps transmissions.
A step further, many local inter-neuronal connections in the cortex are realized  through electrical gap 
junctions \cite{galarreta-hestin:01}, this being predominant between cells of the same sub-population \cite{amitai-etal:02}. 
In such a case the inter-neuron delays are much smaller, but still measurable, since the transmission is mainly due to the spike potential raise, with a time constant 
of about $0.1-0.2 ms$ (see \cite{lewis-rinzel:03} for a discussion about the electrical transmission in this case). 
Then a reasonable assumption is to consider that local electrical connections are delayed  by $d t > \simeq 0.1 ms$.

Gap junctions delays are much smaller ($d t > \simeq 10 \mu s$) but still non negligible \cite{lewis-rinzel:03,koch:99b}.

} \end{quotation}

The [C4] constraint is less obvious. The idea is that, in the absence of any input (isolated neuron), 
the potential decreases towards a resting potential and the neuron cannot fire anymore.
This is true for usual deterministic models, except for singular internal currents\footnote{This is easy to illustrate considering a LIF model, 
where $g$ and $i$ are constant: 
 \\ \centerline{$\left\{ \begin{array}{rcl} C \, \frac{dV}{dt} + g \, V = i, \\ V(t_0) = V_0, V(t_1) = \theta \\ \end{array} \right.
 \Rightarrow t_1 = t_0 + \frac{C}{g} \, \log\left(\frac{i - g \, V_0}{i - g \, \theta}\right) \mbox{ with } i > g \, \theta > g \, V_0.$} \\
If the internal current verify: $i > g \, (\theta - V_0 \, e^{-D \, g / C} ) / (1 - e^{-D \, g / C})$, [C4] is verified. \\ 
Since $C/g \simeq 1 \cdots 10 ms$, thus $e^{-D \, g / C} << 10^4$, it is sufficient to get $i > (1 + 10^{-4}) \, g \, \theta$, 
i.e.  a very small amount above $g \, \theta$. It is thus a reasonable numerical assumption to assume that $D$ is bounded. 
However, if $i \rightarrow g \, \theta$, the firing period becomes unbounded, yielding a spurious event (which can affect the whole dynamic)
at an unbounded instant. This is a singular case.}.
This behaviour seems realistic for cortical neurons, but likely not for all neurons in the brain \cite{pare-bouhassira-etal:90,macormick-bal:97}.

\begin{quotation} { \subsubsection*{Spike time upper-bound evaluation}

At the simulation level, [C4] is easily violated for deterministic neural models with constant internal current, able to integrate during an unbounded period of time, or with maintained sub-threshold oscillations. But this singular condition is easy to check and to avoid, and a maximal spontaneous firing period can be derived. 
Synaptic conductance based models \cite{destexhe:97} and spike response models \cite{gerstner-kistler:02b} usually omit this constant current and their intrinsic ``leak'' guaranties that [C4] is not violated.
On the contrary, with stochastic models, [C4] might be reconsidered, since there is always a ``chance'' to fire a spike,
with a decreasing probability as time increases.

At the biological level\footnote{We are especially thankful to Dr. Thierry Bal, for a scientific discussion on this subject.}, 
in vitro, a regularly spiking cortical pyramidal neuron, without synaptic input, remains silent since its membrane potential is close to the 
resting potential \cite{koch:99b}. In vivo, in the cortex, current observations show that a neuron is always firing \cite{dayan-abbott:01} (unless it is dead). 
This is due to the large amount of neuromodulators, inducing depolarization and a membrane potential close to the firing threshold. \
However, this differs from [C4], where isolated neurons are considered.
On the contrary, thalamic neurons can fire spontaneously after a long resting period \cite{pare-bouhassira-etal:90}.
Even in vitro, their internal currents such as IT (low threshold transient Ca2+ current) or IH (hyper-polarization-activated cation current) can induce
spikes (due to oscillatory behaviors) \cite{macormick-bal:97}.

} \end{quotation}

As discussed in details in \cite{cessac-vieville:08}, the fact whether the constraint [C4] is verified or not completely changes the nature of the dynamics. 
In the latter case, a neuron can remain silent a very long range of time, and then suddenly fire, inducing a complete change in the further state of the system. 
We distinguish situations with and without [C4] in the sequel. 
 
Considering C[1-3] and optionally [C4], let us now review the related consequences regarding modelling and simulation. 

\paragraph{Simulation of time-constrained networks.}

The event-based simulation of spiking neuron networks (see e.g. \cite{brette-rudolph-etal:07} for a review) is strongly simplified by the fact that, 
thanks to [C2] and [C4] spike times and precisions are bounded, while thanks to [C3] spiking can not generate causal paradoxes. 
Here the specification allows to use ``histogram based'' methods\footnote{Source code available at {\tt http://enas.gforge.inria.fr}.},
with a small $O(1)$ complexity \cite{cessac-rochel-etal:09}.

Furthermore, the simulation core is minimal (a 10Kb C++ source code), using a ${\cal O}(D/dt+N)$ buffer size and about ${\cal O}(1+C+\epsilon/dt) \simeq 10-50$ 
operations/spike ({ $>10^6$ spike/sec on a laptop}), for a size $N$ network with $C$ connections in average.

\section{The maximal amount of information}\label{info-amount}

Considering [C1-2], given a network of $N$ spiking neurons observed during a finite period $[0, T]$, 
the number of possible spikes is obviously limited by the refractory period $r$. 
Furthermore, the information contained in all spike times is strictly bounded, since two spike occurrences in a $\delta t$ window are not distinguishable,
and $\delta t < r$.

A rather simple reasoning yields 
a rough {\em upper bound for the amount of information}:
\\ \centerline{$N \, \frac{T}{r} \, \log_2\left(\frac{T}{\delta t}\right)$ bits during $T$ seconds}
Taking the biological values into account, a straightforward numerical derivation leads to about $1 Kbits/neuron$, for $T \gg \delta t$.

\begin{quotation} { \subsubsection*{Information upper bound evaluation}

Let us consider a given neuron (the index number is omitted) and its first spike time $t^1$.
The next spike firing of the neuron, (i) either occurs no later than $t^1 + \delta t$ thus at a time not distinguishable from $t^1$ by an observer,
(ii) or occurs at least $\delta t$ later. 
In order to be meaningful, spikes must thus occur in distinct temporal boxes of width $\delta t$, the precise location of the box being fixed by the first time 
occurrence, as schematized in Fig.~\ref{box-spikes}.
Since there is a refractory period $r > \delta t$ the second and next spikes will never be mixed with their predecessors  
but are going to be subject to the same limitation.

\begin{figure}[h]  
 \begin{center} \includegraphics[width=8cm,height=3cm]{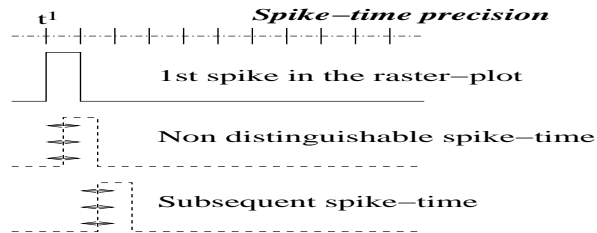} \end{center}
 \caption{\small \label{box-spikes} Evaluating the information in a set of spike times. See text for details.} 
\end{figure}

As a consequence, no more than one spike every $r$ milliseconds can be introduced in this temporal histogram of $\delta t$ box width,
as illustrated in Fig.~\ref{box-spikes}. 
In a $[0, T]$ time range, there are $T/\delta t$ choices for the first spike, less than $T/\delta t - 1$ for the second etc... 
This means that for the $T/r$ maximal number of spikes, they are less than $\left(T/\delta t\right)^{T/r}$ choices. 

Assuming, as a maximal case, that each neuron is independent, we obtain the proposed bound.

} \end{quotation}

This is a rough upper bound that does not take into account constraints imposed by the dynamics at the network level. These constraints further reduce the available information.
In fact, the dynamics of a given network does constraint very much the possible spike trains, and the real entropy may be lower, or even strongly lower, than this bound.

In the particular case of fast-brain mechanisms, where only ``the first spikes matter'' \cite{thorpe-fabre-thorpe:01}, this amount of information is not related to the 
permutations between neuron spikes, i.e. of order of $o(\log(N!)) = N \, \log(N)$  but simply proportional to $N$, in coherence to what is found in \cite{vieville-crahay:04}.

The latter bound is coherent with several results presented in \cite{rieke-etal:96} where the authors consider firing rates and use entropy as information measure. 
For instance, considering a timing precision of $0.1 - 1 ms$ as set here, the authors obtain an information rate bounded around $500 bits/s$ for a neural receptor. 
This number has the same order of magnitude, as obtained by the previous general bound. But the network dynamics itself introduces more specific constraints,
thus  yielding an information rate lower than predicted by the previous bound. 
However, we see here that effective information rate is not an order of magnitude lower: In practice, the dynamics looks lke rich enough to maintain a high information rate.

This information bound is not a bad, but a good news. The result means that different informations are necessarily represented by distinguishable spiking patterns. 
In other words there is a well-defined margin between two different information representations.
The neuronal coding with large margins is discussed in \cite{vieville-crahay:04}, and may explain the surprisingly impressive performance of fast brain categorization. 
This corresponds to introduce an incompressible margin between two informations, which guaranties a robust coding.

\section{Dynamics of time-constrained networks} \label{about-dynamics}

A step further, taking [C1-3] into account, allows us to ``discretize'' the spike train sequences. 
A \textit{raster} is formally defined as follows:
To each neuron of index $i$ a binary variable ${\bf \omega}_i(k) \in \{0, 1\}$ is associated such that the neuron fires 
during the $k$-th sampling period if and only if ${\bf \omega}_i(k) = 1$ and is silent otherwise.
The sampling period is taken smaller than $r$, $\delta t$ and $d t$. 
Smaller than $r$ in order to have either 1 or 0 spike during a sampling period.
Smaller than $\delta t$ in order that the sampling does not impair the spike-time precision.
Smaller than $d t$ since, in a discrete time system, the information if propagated from one sampling period to another through recurrence relations.

In simple models such as basic leaky integrate and fire (LIF) or integrate and fire neuron models with conductance synapses and constant external current (gIF), a full characterization of the network dynamics
can be derived from such a discretization. For these two neuron models, it has been shown that \cite{cessac:08,cessac-vieville:08}: 

\begin{itemize}
\item {[P1] \em The raster is generically\footnote{Considering a basic leaky integrate and fire neuron network the result is true except for a negligible set of parameters. Considering an integrate and fire neuron model with conductance synapses the result is true unless the trajectory accumulates 
on the threshold from below.}
periodic, but, depending on parameters such as constant external current or synaptic weights,
periods can be larger than any accessible computational time;}
\item {[P2] \em 
There is a one-to-one correspondence between orbits\footnote{Here we consider orbits, i.e. infinite trajectories, thus consider this deterministic system, with constant input, in its asymptotic stage.} and rasters (i.e. a raster provides a symbolic coding for the network dynamics).}
\end{itemize}

Note that [P1] and [P2] are properties of usual integrate and fire network models with constant parameters (weights, delays, etc...).

The fact [P1] gives way to clearly understand to which extends spike trains can code information: Periodic orbits give the code. 
When the parameters vary, the orbits change accordingly but are still periodic (with possibly very large periods).

The fact [P2] means that, in the LIF and gIF cases, 
the raster is a ``symbolic coding'' in the sense that no information is lost by considering the spike times instead of the membrane potential variations.

\begin{figure}[htb]  
 \begin{center} \includegraphics[width=8cm,height=4cm]{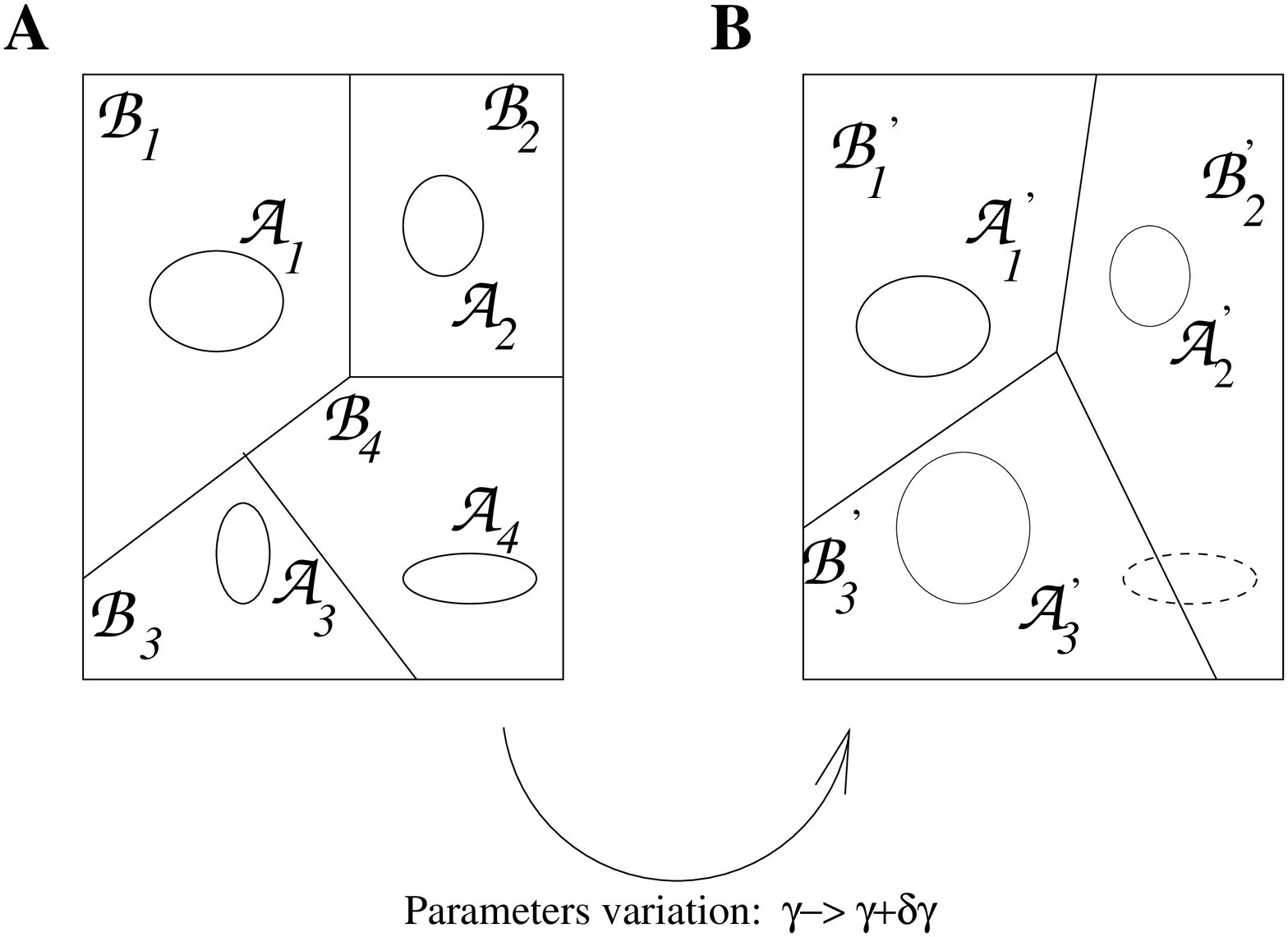} \end{center} 
 \caption{\small \label{attractors} Describing the basins of attraction of the dynamic landscape, for deterministic time-constrained networks.
[{\bf A}] The phase space (in other words the space of the network states) is partitioned into bounded domains ${\cal B}_l$ 
and for each initial condition in ${\cal B}_l$ the initial trajectory is attracted, not towards a fixed point (as in Hopfield networks with asynchron dynamics), 
but towards a periodic orbit ${\cal A}_l$.
[{\bf B}] If the parameters (external input, weights) change, the landscape is modified and several phenomena can occur: 
change in the attractor's shapes, number of attractors, as for ${\cal A}_3$ in this example;
A point belonging to ${\cal A}_4$ in Fig.\ref{attractors} A, can, after modification of the parameters, converge
 either to attractor ${\cal A}'_2$ or ${\cal A}'_3$.}
\end{figure}

Both facts also allow one to deeply understand the network dynamics: Fig.~\ref{attractors} sketches out some aspects, illustrating the global behavior of the system
and illustrating that attractors are generically stable period orbits.
More precisely, the dynamics is piece-wise continuous, i.e. continuous expect when a spike is fired. 
The dynamics is locally contracting. 
Furtermore, after {\em each neuron has fired once the dynamics is no longer dependent on the initial conditions}. 
Nevertheless, when the membrane potential is close to the threshold, a small perturbation may induce drastic changes in the dynamics, while it is otherwise damped.
This behaviour corresponds to a notion of ``edge of chaos'' which is precisely defined
within this framework \cite{cessac:08,cessac-vieville:08}, although this definition differs from the usual notion of chaos in differentiable systems (the terminology ``stable chaos'' has been proposed
by \cite{politi-torcini:09}).

\subsubsection*{Remarks}~\vspace{-0.5cm}

Time is discretized, but without any constraint about the ``sampling period''. The [P1] and [P2] results hold at any finite precision. 
However, to which extends the period of the periodic orbits does not depend on the sampling period, providing the sampling period is small enough, 
or more generally periodic orbits dependence with respect to the sampling period is still an open issue.

In order to understand [P1], it might be important to discuss how ``obvious'' it is. 
Time is discretized. If the membrane potential would have been discretized also, this would have reduced to a finite state system. 
In that case, only fixed points and periodic orbits could occur and the result would have been obvious. 
As a consequence, [P1] reads: \textit{Even if the neuron state is described by continuous values, orbits are still generically periodic.}

In a conductance based model, with the additional constraint that conductances depend on previous spikes within a finite horizon, 
it appears that [P1] still holds, although this is intuitively less obvious \cite{cessac-vieville:08}.

To which extends such a ``canonical situation'' is still true for more complex models is an open question.
We can easily conjecture that [P1] is a model limitation for all integrate and fire models, 
providing they are defined with an instantaneous reset to a constant value. The question is still open for SRM models.

The [P2] statement can be explained as follows.
 Changing the initial value of the membrane potential, one may expect some variability in the dynamics. 
But due the reset, close-by distinct trajectories can be collapsed onto the same trajectory,
after a finite time. As a result, the membrane potential evolution
 then depends only on the previous spike times, instead of the previous membrane potential values
\cite{cessac:08}.

Since periods exhibited by integrate and fire models can be arbitrary large, depending on parameters such as synaptic weights, it is likely
that rasters produced by these models can approach rasters produced by more realistic models such as Hodgkin-Huxley, for a finite horizon.  
However this suggestion is a conjecture only.
This property is reminiscent of the shadowing lemma of dynamical systems theory
\cite{katok-hasselblatt:98} stating that chaotic orbits produced by a uniformly
hyperbolic system can be approached arbitrary close by periodic orbits.

\section{Neural coding and spike train metrics} \label{neural-coding}

In a biological as well as a computational context, the analysis of experimental or simulation data often requires a comparison between two or several spike trains. Either the spike trains concern a given neuron and result from several repetitions of a same experiment, or the spike trains have been generated by different neurons during a given time range, in a unique experiment. In both cases, the idea is to look for invariants, or differences, in the underlying neural code. In the present section and the next two, we study the relation between neural coding and different spike train metrics.

 As an illustrative example, let us consider the temporal order coding scheme \cite{gautrais-thorpe:98,thorpe-fabre-thorpe:01} (i.e. rank coding): Only the order of the events matters, not their specific time values. Two spike trains ${\cal F}_1$, ${\cal F}_2$ with the same event ordering correspond to the same code. This assertion defines an ``equivalence relation'' which structures the set of all the spike trains into a partition: every spike trains in a same equivalence class correspond to the same ``code''.

 Similar definitions can be given for other coding methods. For instance, rate coding means that all spike trains with the same frequency are in the same equivalence class, irrespective of their phase. 

Let us now reconsider the question of neural coding under the light of the time constraints discussed in previous sections.
 The fact that spike time precision is not unbounded leads to many indistinguishable orderings. This does not change the rank coding concept, while the partition is now coarser:
 Trains with two spikes occuring at indistinguishable times are in the same equivalence class. 

 Let us now introduce the notion of spike train metric. The basic idea consists of defining
 a ``distance''
$d(.)$, such that $d({\cal F}_1,{\cal F}_2)=0$
 if ${\cal F}_1$ and ${\cal F}_2$ correspond to the same code, and $1$ otherwise. 

 A step further, how can we capture the fact that, e.g. for rank coding, two spike times with a difference ``about'' $\delta t$ are ``almost'' indistinguishable ? The natural idea is to use a 
``quantitative'' distance instead of a discrete distance (i.e. with binary 0/1 values): 
Two spike trains correspond exactly to the same neural code if the distance is zero
and the distance increases with the difference between the trains.

 This is the idea we wanted to highlight here. This proposal is not a mathematical ``axiomatic'', but a simple {\em modelling choice}. The principle is far for being new, but rather surprisingly it has not been explicited at this level of simplicity. In order to see the interest of the idea, let us briefly review the main classes of spike train metrics.

 As reviewed in details in \cite{schrauwen:07,victor:05} spike trains metrics can be categorized in three classes:
\\ -0- ``Bin'' metrics, based on grouping spikes into bins (e.g. rate coding metrics): Not discussed here.
\\ - I -  Convolution metrics, including the raster-plot metric: Discussed in Section~\ref{continous-metrics}.
\\ -II- Spike time metrics, such as alignment distances \cite{victor-purpura:96}: Discussed in Section~\ref{alignment-metrics}.

\section{Using convolution metrics to link spike trains and continous signals} \label{continous-metrics}

\paragraph{Linear representation.} A large class of metrics is defined through the choice of a convolution kernel $K_i$ applied to a spike train function written $\rho_i(t) = \sum_{t_i^n \in {\cal F}_i} \delta(t - t_i^n)$, where $\delta(.)$ is the Dirac distribution. For a given spike train ${\cal F}_i$, the equation is: 
$$s_i(t) = \sum_{t_i^n \in {\cal F}_i}  K_i(t - t_i^n) = K_i * \rho_i (t) \  \in [0, 1],$$
The signal $s_i$ is easily normalized between $0$ (no spike) and, say, $1$ (burst mode at the maximal frequency). 

The distance between two spike trains is then defined by applying some $L^p$ norm to the continuous signal ${\bf s} = (\cdots, s_i, \cdots)$, at the network level.
The ``code'' here corresponds to the linear representation metric: the codes are similar if the related continuous signals are similar.
It allows us to link spike trains with a continuous signal ${\bf s}$.

\begin{figure}[htb]  
 \begin{center} {\small \begin{tabular}{lll}
{[A]}   & \parbox{2cm}{\includegraphics[width=2.5cm]{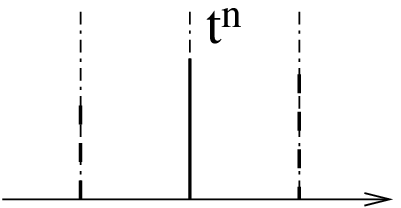}}  & $\delta(t - t_i^n)$ \\
{[B]}  & \parbox{2cm}{\includegraphics[width=2.5cm]{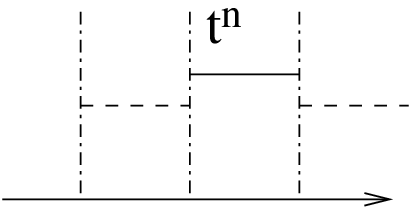}}  & $\chi_{[t_i^{n-1},t_i^n[}(t) \, \frac{r}{t_i^n - t_i^{n-1}}$ \\
{[C]} & \parbox{2cm}{\includegraphics[width=2.5cm]{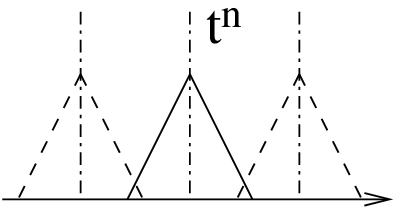}}  & $\mbox{max}\left(0, \frac{r - |t - t_i^n|}{r}\right)$ \\
{[D]}  & \parbox{2cm}{\includegraphics[width=2.5cm]{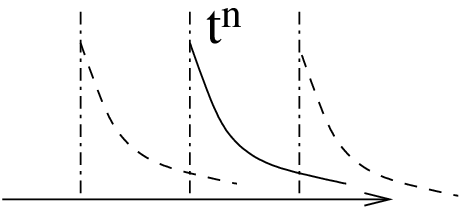}} & $(1-e^\frac{-r}{\tau}) \chi_{[0,\infty[}(t - t_i^n) \, e^\frac{-(t - t_i^n)}{\tau}$ \\
\end{tabular}} \end{center}
\caption{\small \label{kernel} A few examples of spike train convolution: [A] The spike train itself, [B] A causal local frequency measure estimation (writing $\chi$ the indicatrix function), [C] A non-causal spike density, uniformly equal to 1 in burst mode, [D] A normalized causal exponential profile, parameterized by a decay time $\tau$. 
Evoked post-synaptic potential profiles are nothing but such causal convolution (using e.g. double-exponential kernels to capture the synaptic time-constant (weak delay) and potential decay). 
Similarly spike-trains representations using Fourier or Wavelet Transforms are intrinsically related to such convolutions.}
\end{figure}

The so-called ``kernel methods'' based on the Mercer theorem \cite{schrauwen:07} are in direct links with the linear representation since they are defined, 
as scalar products, writing:
$$k({\cal F}, {\cal F}') = \sum_i \sum_{n, m} \hat{K}_i(t_i^n - t_i^{'m}) = 
\int_t s_i(t) \, s'_i(t)dt,$$
with direct correspondences for usual kernels with linear convolutions, e.g.:
$$\scriptsize \begin{tabular}{|l|c|c|c|} \hline & Triangular   & Exponential  & Gaussian 
\\ \hline
 $K_i(t)$ & $\sqrt{\frac{\lambda}{2}}               \, {\cal H}\left(t \, \left(\frac{2}{\lambda} - t\right)\right)$
          & $\sqrt{2 \, \lambda}                    \, {\cal H}(t) \, e^{-\lambda \, t}$
          & $\sqrt{\frac{2 \, \lambda}{\sqrt{\pi}}}         \, e^{-2 \, \lambda^2 \, t^2}$
\\ \hline
 $\hat{K}_i(d)$ & $\mbox{max}\left(1 - \frac{\lambda}{2} \, |d|, 0 \right)$ 
                & $e^{-\lambda \, |d|}$ 
                & $e^{-\lambda^2 \, d^2}$ 
\\ \hline
\end{tabular}
$$
where ${\cal H}$ is the Heaviside function.
Distances based on inter-spike intervals are also included, as developed in e.g. \cite{kreuz-haas:07}.

Non static kernels of the form $K_i^t(t - t_i^n)$ (i.e. depending on $t$) can also be used (clock-dependent coding, raster, 1st spike coding, ...),
while non-linear Volterra series are useful for representing ``higher order'' phenomena (see e.g. \cite{rieke-etal:96}). 

These linear representations not only provide with tools to compare different spike trains, but allows one to better understand the link between continuous signals and spike times.
For instance \cite{dayan-abbott:01,maass:97}, writing $s(t) = \sum_i \lambda_i s_i(t)$ is a mean to define some network readout to link spiking networks to ``analog'' sensory-motor tasks.
Let us illustrate this aspect by the following results. 

\paragraph{Kernel identification.} 

Given a causal signal $\bar{s}_i$ generated by a spike train ${\cal F}_i$ at the unit level, the problem of identifying the related kernel 
is formally solved by the following paradigm:
$$\mbox{min}_{K_i} \int_{t>0} |{s}_i(t) - \bar{s}_i(t)|^2 \, dt \equiv \int_\lambda |K_i(\lambda) \, \rho_i(\lambda) - \bar{s}_i(\lambda)|^2 \, d\lambda,
$$
using the Laplace transform Parseval theorem (here, $\lambda$ is the Laplace domain variable), thus:
$$K_i(\lambda) = \left[ \bar{s}_i(\lambda) \, \rho_i(\lambda)^T \right] \, \left[ \rho_i(\lambda) \, \rho_i(\lambda)^T \right]^{-1}
$$
i.e. the spike train cross-correlation versus auto-correlation ratio. Non-causal estimation would consider the Fourier transform. 
This setting corresponds to several identification methods \cite{dayan-abbott:01,schrauwen:07}.

The paradigm is to be used, for instance, for identifying the average synaptic response profile from the observation of the input spike train and synaptic evoked potential output.
Given the observation of a spike train function $\rho_i$ and the related response $\bar{s}_i$ the previous formula allows one to estimate the related kernel.

\begin{figure}[htb]  
 \centerline{\begin{tabular}{ccc} 
  before & after \\
  \includegraphics[width=5cm,height=3cm]{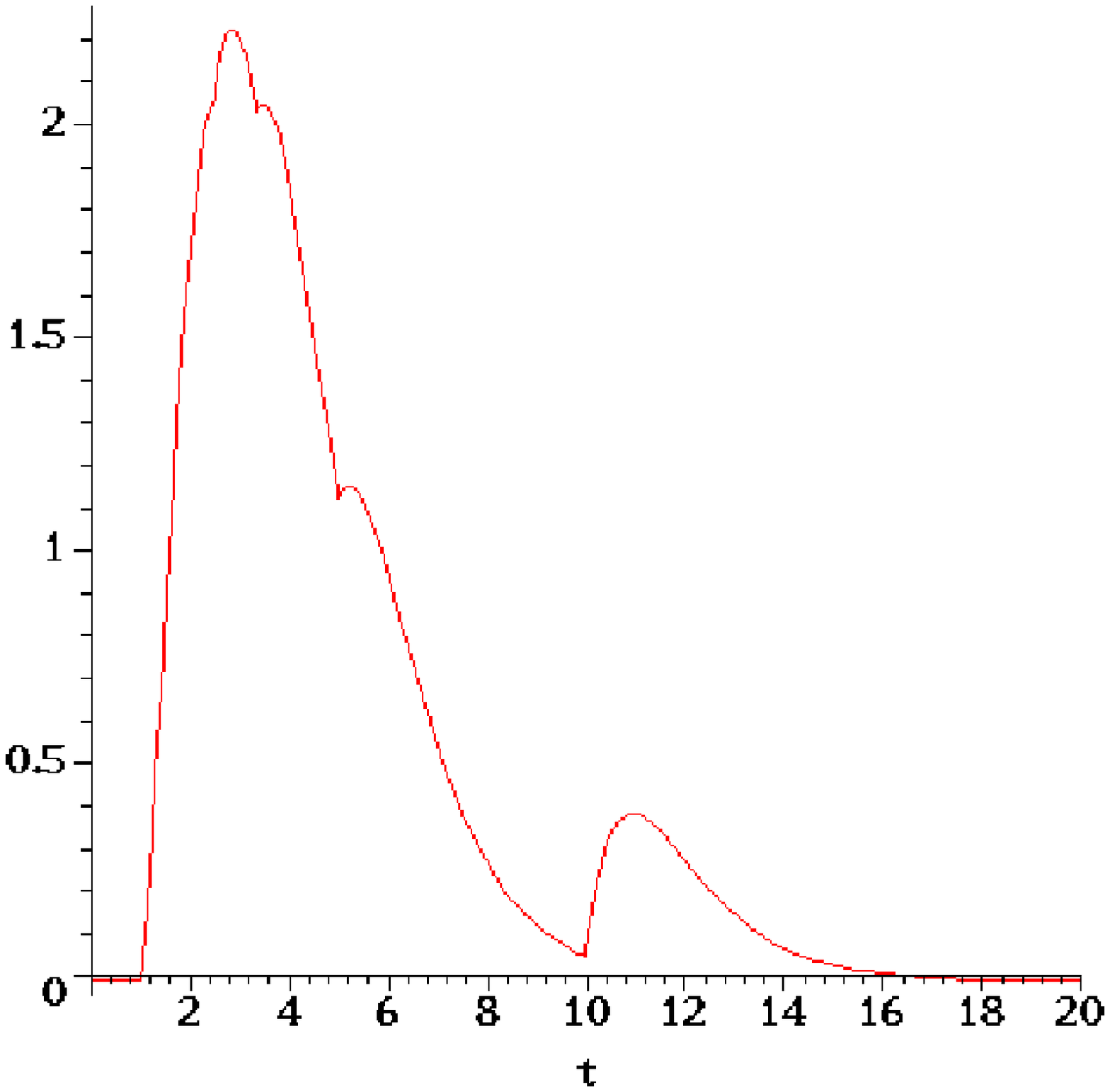} & 
  \includegraphics[width=5cm,height=3cm]{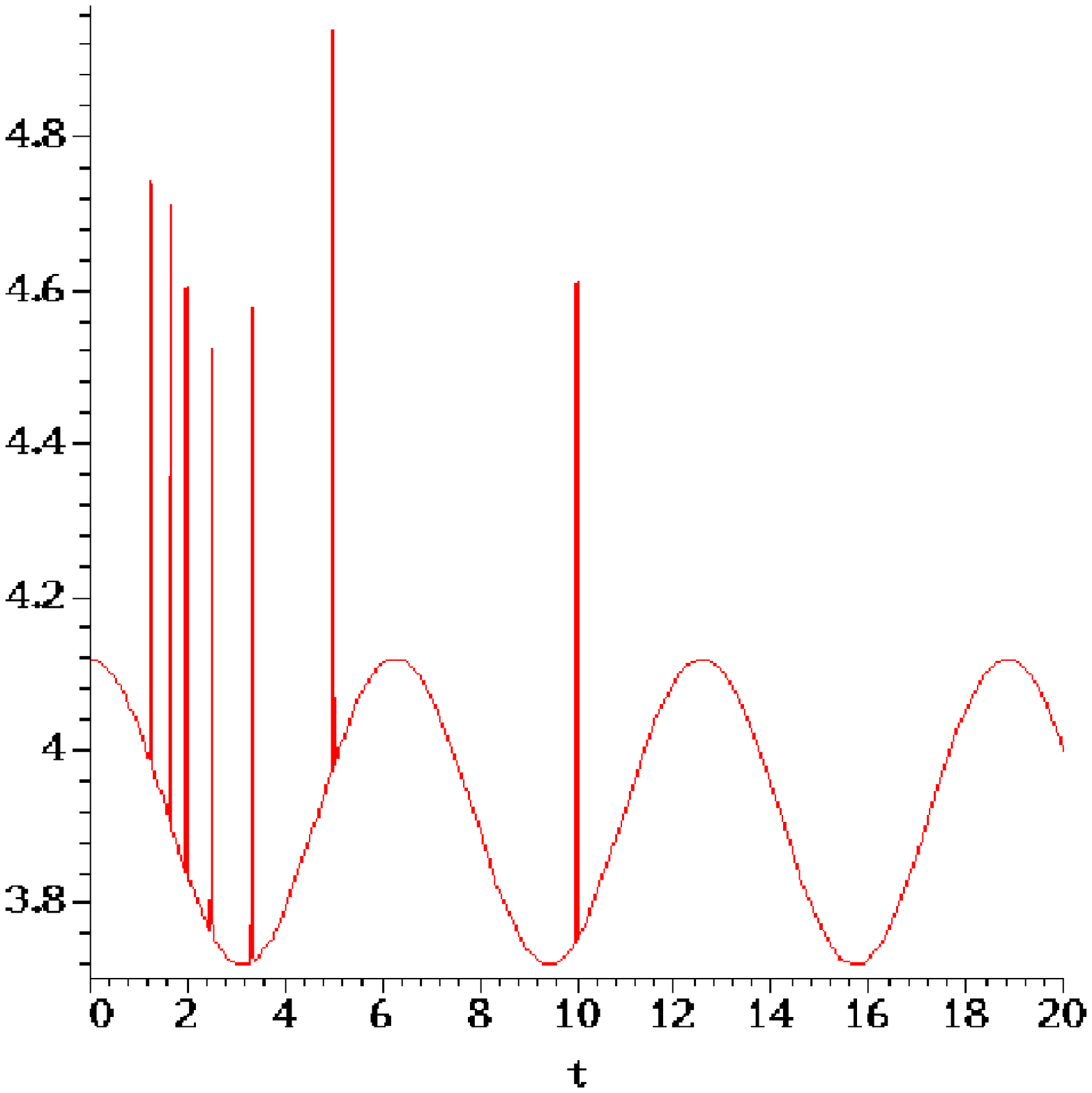}  \\
 \end{tabular}}
\caption{\small \label{deconv} A small experiment of spike deconvolution. 
On the left the signal is the convolution of a spike-train using an $\alpha(t) = t / \tau e^{-t/ \tau}$ profile, 
with addition of noise and of a spurious sinusoid which has been added as an outlier to the signal. 
Spikes are not ``visible'' in the sense that they do not correspond to maxima of the signal because the spike responses are mixed.
On the right the deconvolution is shown: the outlier is amplified, but spikes clearly emerge from the signal.}
\end{figure}

\paragraph{Spike deconvolution.} A step further, if we know the convolution kernel $K_i$, it is obvious to formally write $\rho_i = L_i * s_i$ with 
$L_i = {F}^{-1}\left[\frac{1}{{F}\left[K_i\right]}\right],$
writing ${F}$ the Fourier transform e.g.: 
$$\begin{array}{ll}
     K_i(t) =                 e^{-\frac{t}{T}} & (L_i * s_i) (t) = \frac{1}{\tau} \, s(t) + s'(t) \\
     K_i(t) =  \frac{t}{T} \, e^{-\frac{t}{T}} & (L_i * s_i) (t) = \frac{1}{\tau^2} \, s(t) + \frac{2}{\tau} \, s'(t) + s''(t) \\
  \end{array},$$
well defined and allowing one to reconstruct the spike-train from the continuous signal 
as illustrated in Fig.~\ref{deconv}.

The good news is that the inverse convolution filters $L_i$ are not singular so that the deconvolution is well-defined and  in explicit form. 
However, this requires the use of derivative filters, known as sensible to noise. 
Unpublished numerical investigations have shown that as soon as the error on the kernel profiles is higher than $10-20\%$, several spikes are lost in the deconvolution.

\paragraph{Signal reconstruction.} In order to further understand the power of representation of spike trains \cite{lazar:05} has generalized the well-known Shanon's 
theorem, as follows: A frequency range $[-\Omega, \Omega]$ signal is entirely defined by irregular sampling values $s_i^n$ at spike times $t_i^n$ 
$$s_i(t) = \sum_n K_i^n(t - t_i^n)$$
with
$$K_i^n(t) = s_i^n \, \frac{\sin(\Omega t)}{\pi \, t},$$
provided that $max_n d_i^n \leq \frac{\pi}{\Omega}$. 

This supplies an explicit signal ``decoding'', 
since given any signal $s$ it provides an explicit formula to represent this signal by a convolution kernel $K$ and a spike train.

\paragraph{Raster metrics.} A step further, it is easy to see that representing the spike time by a raster  corresponds to a non-static convolution kernel. 
A given raster can be represented by a real number in $[0 , 1[$, the binary representation of its decimal part being the spike train itself. 
Using this representation, a useful related metric is of the form, for $\theta \in ]0, 1[$:
$$d_\theta({\bf \omega}, {\bf \omega}') = \theta^T, T = \mbox{argmax}_t \;  {\bf \omega}^t = {\bf \omega}'^t,
$$
thus capturing the fact that two rasters are equal up to a certain rank. Such metrics can be applied to analyze the dynamics of spiking networks 
and they are typically used in the context of symbolic coding in dynamical systems theory \cite{cessac:08,cessac-vieville:08}.

\section{Using alignment metrics to program spiking neuron networks} \label{alignment-metrics}

\paragraph{The original alignment metric.}
The second family of metrics we want to review considers spike times directly \cite{victor-purpura:96,victor:05}. 

The distance between two finite spike trains ${\cal F}$, ${\cal F}'$ is defined in terms of the minimum cost of transforming one spike train into another.
Two kinds of operations are defined:
\begin{itemize}
\item spike insertion or spike deletion, the cost of each operation being set to $1$
\item spike shift, the cost to shift from $t_i^n \in {\cal F}$ to $t_i^{'m} \in {\cal F}'$ being set to $|t_i^n - t_i^{'m}| / \tau$ for a time constant $\tau$.
\end{itemize}

For small $\tau$, the distance approaches the number of non-coincident spikes, since instead of shifting spikes it is cheaper to insert/delete non-coincident spikes, the distance being always bounded by the number of spikes in both trains.

For high $\tau$, the distance basically equals the difference in spike number (rate distance), while for two spike trains with the same number of spikes, 
there is always a time-constant $\tau$ large enough for the distance to be equal to $\sum_n |t_i^n - t_i^{'n}| / \tau$.

Here, two spike times are comparable if they occur within an interval of $2\,\tau$, otherwise they had better to be deleted / inserted.

Although computing such a distance seems subject to a combinatorial complexity, it appears that quadratic algorithms are available (i.e. with a complexity equal to the product of the numbers of spikes). This is due to the fact that, in a minimal path, each spike can be either deleted or shifted once to coincide with a spike in the other spike train. Also, a spike can be inserted only at a time that matches the occurrence of a spike in the other spike train. It allows us 
 to calculate iteratively the minimal distance considering the distance $d_{n,n'}({\cal F}, {\cal F}')$ between a spike train composed of the first $n$ spikes of ${\cal F}$ and the first $n'$ spikes of ${\cal F}'$.


\begin{figure}[htb]  
 \begin{center} \includegraphics[width=7cm,height=1.4cm]{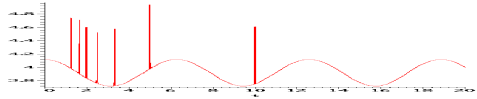} \end{center}
\caption{\small \label{f-to-f} An example of minimal alignment from the upper to the lower spike train, using from top to bottom an insertion, a rightward shift, a leftward shift and a deletion respectively.}
\end{figure}

When considering spike trains with more than one unit, an approach consists to sum the distances for each alignment unit-to-unit. Another point of view is to consider that a spike can ``jump'', with some cost, from one unit in ${\cal F}$ to another unit in ${\cal F}'$. The related algorithmic complexity is no more quadratic but to the power of the number of units \cite{aronov:03}.

This family of metrics include aligments not only on spike times, but also on inter-spike intervals, or metrics which are sensitive to patterns of spikes, etc... They have been fruitfully applied to a variety of neural systems, in order to characterize neuronal variability and coding \cite{victor:05}. For instance, in a set of neurons, that act as coincidence detectors, with integration time (or temporal resolution) $\tau$, spike trains will have similar postsynaptic effects if they are similar w.r.t. this metric.

\paragraph{Generalization of the alignment metric.}

Let us remark, here, that the previous metric can be generalized as follows:

 - [causality] At a given time, the cost of the alignment of previous spikes decreases with the obsolescence of the spike, say, 
with an exponential profile parameterized by a time-constant $\tau'$. 
At the infinity limit for $\tau'$, the original alignment metric is retrieved.

 - [non-linearity] The cost of a shift is not necessarily a linear function of $\frac{|t_i^n - t_i^{'n}|}{\tau}$, as in the original metric,
but any suitable non-linear function $\phi\left(\frac{|t_i^n - t_i^{'n}|}{\tau}\right)$. 

For instance, we may choose a small quadratic profile when lower that the time precision 
(accounting for additive noise, but implementing the fact that spike time differences are negligible), and then, a linear profile.

This leads to an iterative definition of the previous distance $d_{n,n'}$ :
$$
\small d_{n,n'} = \mbox{min}\left(\begin{array}{l} e^{-\frac{t_i^n - t_i^{n-1}}{\tau'}} \, d_{n-1,n'} + 1, \\ e^{-\frac{t_i^{'n} - t_i^{'n-1}}{\tau'}} \, d_{n,n'-1} + 1, \\ e^{-\frac{\max(t_i^n, t_i^{'n}) - \min (t_i^{n-1}, t_i^{'n-1})}{\tau'}} \, d_{n-1,n'-1} + \phi\left(\frac{|t_i^n - t_i^{'n}|}{\tau}\right)\end{array}\right),$$
with, e.g., $\phi(d) = \min\left(d,(d\,\tau/\delta t)^2\right)$, again implementable in quadratic time. 
It corresponds to the original alignment metric iff $\phi()$ is the identity function and $\tau' = +\infty$, still calculable with a quadratic complexity.

This modified version of the metric illustrates how versatile is this class of distances for representing the differences between spike trains.


\paragraph{Weight training from spike times.} As a formal application, let us consider a Spike Response Model neuron \cite{gerstner-kistler:02} of the form:
~\\
\centerline{\small $V_i(t) = \nu(t - t_i^{n-1}) + \sum_{jm} w_{ij} \, \alpha(t - t_j^m)$ \ \ \ \mbox{for} \  $t_i^{n-1} < t \leq t_i^n$,}
the spike time being defined by $V_i(t_i^n) = \theta$, where $\theta$ is the spiking threshold. 

Previous metrics on spike times give way us to optimize the neural weights in order to tune spike times, deriving, e.g., rules of the form:
~\\
\centerline{$ \Delta w_{ij} \equiv \sum_n (t_i^n - \bar{t}_i^n) \, \frac{\partial V_i}{\partial w_{ij}}(t_i^n) \left/ \frac{\partial V_i}{\partial t_i^n}(t_i^n) \right.$}

Such mechanisms of optimization are also applicable to time-constants, delays or thresholds. 
It appears that this method cannot be easily used in practice, since the equation is numerically unstable \cite{schrauwen:07}.
However, using spike train metrics leads to the formalization of such adaptation rules, in order to ``compute with spikes''.

Let us further develop this point now.





\section{Implementing spiking neuron networks}

\subsubsection*{Spiking neuron network models}

In biological context, spiking neuron networks are useful for modelling different areas identified in the brain by neurophysiological experiments and to validate, or invalidate, hypotheses made on their possible functional interactions. For instance, in the ANR MAPS project, interactions between Superior Colliculus (SC), Excitatory Burst Neurons (EBN), central Reticulate Mesencephalic Formation (cMRF), OmniPause Neurons (OPN) and MotoNeurons (MN) are modelized by large size spiking neuron networks in order to explain the control mechanisms of ocular saccades (work in progress).

In computational context, spiking neuron networks are mainly implemented through specific network architectures, such as Echo State Networks \cite{jaeger:03} and Liquid Sate Machines \cite{maass-etal:02}, that are called ``reservoir computing'' 
(see \cite{verstraeten-etal:07} for unification of reservoir computing methods at the experimental level). 
In this framework, the reservoir is a network model of neurons (can be linear or sigmoid neurons, but more usually spiking neurons), with a random topology and a sparse connectivity. The reservoir is a recurrent network, with weights than can be either fixed or driven by an unsupervised learning mechanism.
In case of spiking neurons (e.g. in the model of \cite{paugam-moisy-etal:08}), the learning mechanism is a form of synaptic plasticity, usually STDP (Spike-Time-Dependent Plasticity), a temporal Hebbian unsupervised learning rule, biologically inspired.
The output layer of the network (the so-called ``readout neurons'') is driven by a supervised learning rule, generated from any type of classifier or regressor, ranging from a least mean squares rule to sophisticated discriminant or regression algorithms. 
The ease of training and a guaranteed optimality guides the choice of the method.
It appears that simple methods yield good results \cite{verstraeten-etal:07}.
This distinction between a readout layer and an internal reservoir is indeed induced by the fact that only the output of the neuron network activity is constrained, whereas the internal state is not controlled.

\subsubsection*{Calculability of neural networks}

Let us now consider the \textit{calculability} of neuron network models. 
It is known that recurrent neuron networks with frequency rates are universal approximators \cite{schafer-zimmermann:06}, 
as multilayer feed-forward networks are \cite{hornik-etal:89}.
This means that neuron networks are able to simulate dynamical systems, not only to approximate measurable functions on a compact domain, 
as originally stated (see, e.g., \cite{schafer-zimmermann:06} for a detailed introduction on these notions).
Spiking neuron networks have been proved to be also universal approximators \cite{maass:01}.

\subsubsection*{Learning the parameters of a spiking neuron networks}


In biological context, learning is mainly related to 
synaptic plasticity \cite{gerstner-kistler:02,cooper-intrator-etal:04} and STDP 
(see e.g., \cite{toyoizumi-etal:07} for a recent formalization), as far as spiking neuron networks are concerned. This unsupervised learning mechanism is known to reduce the variability of neuron responses
\cite{bohte-mozer:07} and related to the maximization of information transmission \cite{toyoizumi-etal:05} and mutual information \cite{chechik:03}.
It has also other interesting computational properties such as tuning neurons to react as soon as possible to the earliest spikes, or segregate the network response
in two classes depending on the input to be discriminated, and more general structuring such as emergence of orientation selectivity
 \cite{guyonneau-vanrullen-etal:04}.

In the present study, the point of view is quite different: we consider supervised learning while,
since ``each spike may matter'' \cite{guyonneau-vanrullen-etal:04,delorme-perrinet-etal:01}, 
we want not only to statistically reproduce the spiking output, but also to reproduce it exactly.

The motivation to explore this track is twofold. On one hand we want to better understand what can be learned at a theoretical level by spiking neuron networks, 
tuning weights and delays. The key point is the non-learnability of spiking neurons \cite{sima-sgall:05}, 
since it is proved that this problem is NP-complete, when considering the estimation of both weights and delays. 
Here we show that we can ``elude'' this caveat and propose an alternate efficient estimation, inspired by biological models.

We also have to notice, that the same restriction apply not only to simulation but, as far as this model is biologically plausible, also holds at the biological level. 
It is thus an issue to wonder if, in biological neuron networks, delays are really estimated during learning processes, or if a weaker form of weight adaptation, as developed now, is considered.

On the other hand, the computational use of spiking neuron networks in the framework of reservoir computing or beyond \cite{schrauwen:07}, at application levels, 
requires efficient tuning methods not only in ``average'', but in the deterministic case. 
This is the reason why we must consider how to {\em exactly} generate a given spike train.

\subsubsection*{Weak estimation of network parameters}

As pointed out previously, the non-learnability of spiking neurons is known \cite{sima-sgall:05}, i.e. the previous estimation is proved to be NP-complete. 
This means that in order to ``learn'' the proper parameters we have to ``try all possible combinations of delays''. 
This is intuitively due to the fact that each delay has no ``smooth'' effect on the dynamics but may change the whole dynamics in an unpredictable way.

This is the way proposed to elude this NP-complete problem by considering {\em another} estimation problem.
Here we do not estimate {\em one} delay (for each synapse) but consider connection weights at several delays and 
then estimate a balancing of their relative contribution. 
This means that we consider a {\em weak} delay estimation problem.

The alternative approach is to estimate delayed weights, i.e. a quantitative weight value $W_{ijd}$ at each delay $d \in \{1, D\}$, using e.g. a model of the form:
$$V_i[k] = \gamma_i \, V_i[k - 1] \, (1 - Z_i[k - 1]) + \sum_{j = 1}^n \sum_{d = 1}^D W_{ijd} \, Z_j[k - d] + I_{ik}.$$

Obviously, the case where there is a weight $W_{ij}$ with a corresponding delay $d_{ij} \in \{0, D\}$ is a particular case of considering several delayed weights,
since we can write:
\\ \centerline{$W_{ijd} = W_{ij} \, \delta(d - d_{ij})$,}\\
$\delta()$ being the Kronecker symbol in this case.
In other words, with our weaker model, we are still able to estimate a neuron network with adjustable synaptic delays. 

We thus do not restrain the neuron network model by changing the problem, but enlarge it.
In fact, the present estimation provides a smooth approximation of the previous NP-complete problem.

It has been made explicit in \cite{rostro-etal:09} that the parameter estimation of such a neuron network in order to generate a given spike train,
is a Linear (L) problem if the membrane potentials are observed and a Linear Progamming (LP) problem if only  spike times are observed, with a gIF model. 
Such L or LP adjustment mechanisms are distributed and have the same structure as an ``Hebbian'' rule.
A step further, this paradigm is easily generalizable to the design of input-output spike train transformations.
This means that a practical method is available to ``program'' a spiking network, 
i.e. to find a set of parameters allowing us to exactly reproduce the network output, given an input.

\paragraph{Polychronization and limitations of metrics}

A spiking network can polychronize, i.e., exhibit reproducible time-locked but not synchronous firing patterns within $1$ millisecond precision. Polychronization can be viewd as a generalization of the notions of synchronization and synfire chains.
Due to the interplay between the delays and a form synaptic plasticity (can be implemented by way of STDP - see Section\~ref{}), the spiking neurons spontaneously self-organize into groups and generate patterns of stereotypical polychronous activity.

In \cite{izhikevich:06}, it has been shown that the number of co-existing polychronous groups far exceeds the number of neurons in the network, 
resulting in an unprecedented memory capacity of the system.
The author speculates on the significance of polychrony to the theory of neuronal group selection and cognitive neural computations.

In \cite{paugam-moisy-etal:08}, the network processing and the resulting performance is explained by the concept of polychronization,
The model emphasizes that polychronization can be used as a tool for exploiting the computational power of synaptic delays and for monitoring the topology and
activity of a spiking neuron network \cite{martinez-paugam:08}.

Taking such complex aspects of the neural code into account cannot be performed by any available metrics.
New metrics, taking long term interactions into account have to be developed and this is a challenging issue.

\section{Conclusion}

This article has reviewed a set of indisputable facts that could help better understanding to which extend computing and modelling with spiking neuron networks can be biologically plausible and computationally efficient. The links between spike trains and neural coding have been highlighted, with the help of several metrics and under a set of time constraints as hypotheses.

Although probabilistic measures of spike patterns such as correlations \cite{gerstner-kistler:02} or entropy based pseudo-distances (e.g. mutual information) provide a view of spike trains variability which is enriched by the information theory conceptual framework, it may be difficult to estimate them in practice, since such measures are robust only if a large amount of samples is available. 
On the contrary, distances allow to characterize aspects of spike coding, with efficient methods and without this curse of the sampling size.

This review highlights some of these methods and propose to consider that ``choosing a coding'' means ``defining a metric''. 
This point of view provides a synthetic insight of several methods applied to spiking neuron networks. 
To our best knowledge, only polychronization mechanisms are not easily represented with such a tool, 
and it is an interesting issue to study the link between these non-local temporal interactions in neuron networks and the underlying neural code.

Neither ``incredible power of spikes'' nor ``mystery of the [spike based] neural code'' here, but some pragmatical and practical facts to 
better understand to which extend computing and modelling using spiking neuron networks can be useful, and how to implement such networks in a pertinent way.

{\scriptsize \bibliographystyle{apacite} \bibliography{../../../Latex/string,../../../Latex/odyssee,../../../Latex/biblio}}

\begin{thebibliography}{}

\bibitem[\protect\citeauthoryear{%
Amitai%
\ \protect\BOthers{.}}{%
Amitai%
\ \protect\BOthers{.}}{%
{\protect\APACyear{2002}}%
}]{%
amitai-etal:02}%
\APACinsertmetastar{%
amitai-etal:02}%
Amitai, Y.%
, Gibson, J.%
, Beirleiner, M.%
, Patrick, S.%
, Ho, A.%
, B.W.Connors%
\BCBL{}\ \BOthersPeriod{.}%
\unskip\
\newblock
\APACrefYearMonthDay{2002}{}{}.
\newblock
\BBOQ{}\APACrefatitle{The spatial dimensions of electrically coupled networks
  of interneurons in neocortex}{The spatial dimensions of electrically coupled
  networks of interneurons in neocortex}.\BBCQ{}
\newblock
\APACjournalVolNumPages{J. Neurosci.}{22}{}{4142--4152}.
\PrintBackRefs{\CurrentBib}

\bibitem[\protect\citeauthoryear{%
Aronov%
}{%
Aronov%
}{%
{\protect\APACyear{2003}}%
}]{%
aronov:03}%
\APACinsertmetastar{%
aronov:03}%
Aronov, D.%
%
\unskip\
\newblock
\APACrefYearMonthDay{2003}{}{}.
\newblock
\BBOQ{}\APACrefatitle{Fast algorithm for the metric-space analysis of
  simultaneous responses of multiple single neurons}{Fast algorithm for the
  metric-space analysis of simultaneous responses of multiple single
  neurons}.\BBCQ{}
\newblock
\APACjournalVolNumPages{Journal of Neuroscience Methods}{124}{2}{}.
\PrintBackRefs{\CurrentBib}

\bibitem[\protect\citeauthoryear{%
Baudot%
}{%
Baudot%
}{%
{\protect\APACyear{2007}}%
}]{%
baudot-marre-etal:07}%
\APACinsertmetastar{%
baudot-marre-etal:07}%
Baudot, P.%
%
\unskip\
\newblock
\APACrefYear{2007}.
\newblock
\APACrefbtitle{Nature is the code: high temporal precision and low noise in
  V1}{Nature is the code: high temporal precision and low noise in v1}.
\newblock
\BUPhD.
\PrintBackRefs{\CurrentBib}

\bibitem[\protect\citeauthoryear{%
Bohte%
\ \BBA{} Mozer%
}{%
Bohte%
\ \BBA{} Mozer%
}{%
{\protect\APACyear{2007}}%
}]{%
bohte-mozer:07}%
\APACinsertmetastar{%
bohte-mozer:07}%
Bohte, S\BPBI M.%
\BCBT{}\ \BBA{} Mozer, M\BPBI C.%
%
\unskip\
\newblock
\APACrefYearMonthDay{2007}{}{}.
\newblock
\BBOQ{}\APACrefatitle{Reducing the Variability of Neural Responses: A
  Computational Theory of Spike-Timing-Dependent Plasticity.}{Reducing the
  variability of neural responses: A computational theory of
  spike-timing-dependent plasticity.}\BBCQ{}
\newblock
\APACjournalVolNumPages{Neural Computation}{19}{2}{371--403}.
\PrintBackRefs{\CurrentBib}

\bibitem[\protect\citeauthoryear{%
Brette%
\ \BBA{} Gerstner%
}{%
Brette%
\ \BBA{} Gerstner%
}{%
{\protect\APACyear{2005}}%
}]{%
brette-gerstner:05}%
\APACinsertmetastar{%
brette-gerstner:05}%
Brette, R.%
\BCBT{}\ \BBA{} Gerstner, W.%
%
\unskip\
\newblock
\APACrefYearMonthDay{2005}{}{}.
\newblock
\BBOQ{}\APACrefatitle{Adaptive exponential integrate-and-fire model as an
  effective description of neuronal activity}{Adaptive exponential
  integrate-and-fire model as an effective description of neuronal
  activity}.\BBCQ{}
\newblock
\APACjournalVolNumPages{Journal of Neurophysiology}{94}{}{3637--3642}.
\PrintBackRefs{\CurrentBib}

\bibitem[\protect\citeauthoryear{%
Brette%
\ \protect\BOthers{.}}{%
Brette%
\ \protect\BOthers{.}}{%
{\protect\APACyear{2007}}%
}]{%
brette-rudolph-etal:07}%
\APACinsertmetastar{%
brette-rudolph-etal:07}%
Brette, R.%
, Rudolph, M.%
, Carnevale, T.%
, Hines, M.%
, Beeman, D.%
, Bower, J\BPBI M.%
\BCBL{}\ \BOthersPeriod{.}%
\unskip\
\newblock
\APACrefYearMonthDay{2007}{}{}.
\newblock
\BBOQ{}\APACrefatitle{Simulation of networks of spiking neurons: a review of
  tools and strategies}{Simulation of networks of spiking neurons: a review of
  tools and strategies}.\BBCQ{}
\newblock
\APACjournalVolNumPages{Journal of Computational
  Neuroscience}{23}{3}{349--398}.
\PrintBackRefs{\CurrentBib}

\bibitem[\protect\citeauthoryear{%
Burnod%
}{%
Burnod%
}{%
{\protect\APACyear{1993}}%
}]{%
burnod:93}%
\APACinsertmetastar{%
burnod:93}%
Burnod, Y.%
%
\unskip\
\newblock
\APACrefYear{1993}.
\newblock
\APACrefbtitle{An adaptive neural network: the cerebral cortex}{An adaptive
  neural network: the cerebral cortex}.
\newblock
\APACaddressPublisher{}{Masson, Paris}.
\newblock
\APACrefnote{2nd edition}
\PrintBackRefs{\CurrentBib}

\bibitem[\protect\citeauthoryear{%
Camera%
, Giugliano%
, Senn%
\BCBL{}\ \BBA{} Fusi%
}{%
Camera%
\ \protect\BOthers{.}}{%
{\protect\APACyear{2008}}%
{\protect\APACexlab{{\protect\BCnt{1}}}}}]{%
la-camera-etal:08a}%
\APACinsertmetastar{%
la-camera-etal:08a}%
Camera, G\BPBI L.%
, Giugliano, M.%
, Senn, W.%
\BCBL{}\ \BBA{} Fusi, S.%
%
\unskip\
\newblock
\APACrefYearMonthDay{2008{\protect\BCnt{1}}}{}{}.
\newblock
\BBOQ{}\APACrefatitle{The response of cortical neurons to in vivo-like input
  current: theory and experiment}{The response of cortical neurons to in
  vivo-like input current: theory and experiment}.\BBCQ{}
\newblock
\APACjournalVolNumPages{Biological Cybernetics}{99}{4-5}{279-301}.
\PrintBackRefs{\CurrentBib}

\bibitem[\protect\citeauthoryear{%
Camera%
, Giugliano%
, Senn%
\BCBL{}\ \BBA{} Fusi%
}{%
Camera%
\ \protect\BOthers{.}}{%
{\protect\APACyear{2008}}%
{\protect\APACexlab{{\protect\BCnt{2}}}}}]{%
la-camera-etal:08b}%
\APACinsertmetastar{%
la-camera-etal:08b}%
Camera, G\BPBI L.%
, Giugliano, M.%
, Senn, W.%
\BCBL{}\ \BBA{} Fusi, S.%
%
\unskip\
\newblock
\APACrefYearMonthDay{2008{\protect\BCnt{2}}}{}{}.
\newblock
\BBOQ{}\APACrefatitle{The response of cortical neurons to in vivo-like input
  current: theory and experiment: II. Time-varying and spatially distributed
  inputs}{The response of cortical neurons to in vivo-like input current:
  theory and experiment: Ii. time-varying and spatially distributed
  inputs}.\BBCQ{}
\newblock
\APACjournalVolNumPages{Biological Cybernetics}{99}{4-5}{303-318}.
\PrintBackRefs{\CurrentBib}

\bibitem[\protect\citeauthoryear{%
Carandini%
\ \protect\BOthers{.}}{%
Carandini%
\ \protect\BOthers{.}}{%
{\protect\APACyear{2005}}%
}]{%
carandini-demb-etal:05}%
\APACinsertmetastar{%
carandini-demb-etal:05}%
Carandini, M.%
, Demb, J\BPBI B.%
, Mante, V.%
, Tollhurst, D\BPBI J.%
, Dan, Y.%
, Olshausen, B\BPBI A.%
\BCBL{}\ \BOthersPeriod{.}%
\unskip\
\newblock
\APACrefYearMonthDay{2005}{{\APACmonth{11}}}{}.
\newblock
\BBOQ{}\APACrefatitle{Do we know what the early visual system does?}{Do we know
  what the early visual system does?}\BBCQ{}
\newblock
\APACjournalVolNumPages{Journal of Neuroscience}{25}{46}{10577--10597}.
\PrintBackRefs{\CurrentBib}

\bibitem[\protect\citeauthoryear{%
Carandini%
\ \BBA{} Ferster%
}{%
Carandini%
\ \BBA{} Ferster%
}{%
{\protect\APACyear{2000}}%
}]{%
carandini-ferster:00}%
\APACinsertmetastar{%
carandini-ferster:00}%
Carandini, M.%
\BCBT{}\ \BBA{} Ferster, D.%
%
\unskip\
\newblock
\APACrefYearMonthDay{2000}{}{}.
\newblock
\BBOQ{}\APACrefatitle{Membrane Potential and Firing Rate in Cat Primary Visual
  Cortex}{Membrane potential and firing rate in cat primary visual
  cortex}.\BBCQ{}
\newblock
\APACjournalVolNumPages{The Journal of Neuroscience}{20}{1}{470--484}.
\PrintBackRefs{\CurrentBib}

\bibitem[\protect\citeauthoryear{%
Cessac%
}{%
Cessac%
}{%
{\protect\APACyear{2008}}%
}]{%
cessac:08}%
\APACinsertmetastar{%
cessac:08}%
Cessac, B.%
%
\unskip\
\newblock
\APACrefYearMonthDay{2008}{}{}.
\newblock
\BBOQ{}\APACrefatitle{A discrete time neural network model with spiking
  neurons. Rigorous results on the spontaneous dynamics}{A discrete time neural
  network model with spiking neurons. rigorous results on the spontaneous
  dynamics}.\BBCQ{}
\newblock
\APACjournalVolNumPages{J. Math. Biol.}{56}{3}{311-345}.
\PrintBackRefs{\CurrentBib}

\bibitem[\protect\citeauthoryear{%
Cessac%
, Rochel%
\BCBL{}\ \BBA{} Vi\'eville%
}{%
Cessac%
\ \protect\BOthers{.}}{%
{\protect\APACyear{2009}}%
}]{%
cessac-rochel-etal:09}%
\APACinsertmetastar{%
cessac-rochel-etal:09}%
Cessac, B.%
, Rochel, O.%
\BCBL{}\ \BBA{} Vi\'eville, T.%
%
\unskip\
\newblock
\APACrefYearMonthDay{2009}{}{}.
\newblock
\BBOQ{}\APACrefatitle{Introducing numerical bounds to improve event-based
  neural network simulations}{Introducing numerical bounds to improve
  event-based neural network simulations}.\BBCQ{}
\newblock
\APACjournalVolNumPages{Frontiers in neuroscience}{}{}{}.
\newblock
\APACrefnote{submitted}
\PrintBackRefs{\CurrentBib}

\bibitem[\protect\citeauthoryear{%
Cessac%
\ \BBA{} Vi{\'e}ville%
}{%
Cessac%
\ \BBA{} Vi{\'e}ville%
}{%
{\protect\APACyear{2008}}%
}]{%
cessac-vieville:08}%
\APACinsertmetastar{%
cessac-vieville:08}%
Cessac, B.%
\BCBT{}\ \BBA{} Vi{\'e}ville, T.%
%
\unskip\
\newblock
\APACrefYearMonthDay{2008}{jul}{}.
\newblock
\BBOQ{}\APACrefatitle{On Dynamics of Integrate-and-Fire Neural Networks with
  Adaptive Conductances}{On dynamics of integrate-and-fire neural networks with
  adaptive conductances}.\BBCQ{}
\newblock
\APACjournalVolNumPages{Frontiers in neuroscience}{2}{2}{}.
\PrintBackRefs{\CurrentBib}

\bibitem[\protect\citeauthoryear{%
Chechik%
}{%
Chechik%
}{%
{\protect\APACyear{2003}}%
}]{%
chechik:03}%
\APACinsertmetastar{%
chechik:03}%
Chechik, G.%
%
\unskip\
\newblock
\APACrefYearMonthDay{2003}{}{}.
\newblock
\BBOQ{}\APACrefatitle{Spike-Timing-Dependent Plasticity and Relevant Mutual
  Information Maximization}{Spike-timing-dependent plasticity and relevant
  mutual information maximization}.\BBCQ{}
\newblock
\APACjournalVolNumPages{Neural Computation}{15}{7}{1481--1510}.
\PrintBackRefs{\CurrentBib}

\bibitem[\protect\citeauthoryear{%
Cooper%
, Intrator%
, Blais%
\BCBL{}\ \BBA{} Shouval%
}{%
Cooper%
\ \protect\BOthers{.}}{%
{\protect\APACyear{2004}}%
}]{%
cooper-intrator-etal:04}%
\APACinsertmetastar{%
cooper-intrator-etal:04}%
Cooper, L.%
, Intrator, N.%
, Blais, B.%
\BCBL{}\ \BBA{} Shouval, H.%
%
\unskip\
\newblock
\APACrefYear{2004}.
\newblock
\APACrefbtitle{Theory of Cortical Plasticity}{Theory of cortical plasticity}.
\newblock
\APACaddressPublisher{}{World Scientific Publishing}.
\PrintBackRefs{\CurrentBib}

\bibitem[\protect\citeauthoryear{%
Crook%
, Ermentrout%
\BCBL{}\ \BBA{} Bower%
}{%
Crook%
\ \protect\BOthers{.}}{%
{\protect\APACyear{1998}}%
}]{%
crook-ermentrout-etal:98}%
\APACinsertmetastar{%
crook-ermentrout-etal:98}%
Crook, S.%
, Ermentrout, G.%
\BCBL{}\ \BBA{} Bower, J.%
%
\unskip\
\newblock
\APACrefYearMonthDay{1998}{}{}.
\newblock
\BBOQ{}\APACrefatitle{Spike frequency adaptation affects the synchronization
  properties of networks of cortical oscillations}{Spike frequency adaptation
  affects the synchronization properties of networks of cortical
  oscillations}.\BBCQ{}
\newblock
\APACjournalVolNumPages{Neural Computation}{10}{4}{}.
\PrintBackRefs{\CurrentBib}

\bibitem[\protect\citeauthoryear{%
Dayan%
\ \BBA{} Abbott%
}{%
Dayan%
\ \BBA{} Abbott%
}{%
{\protect\APACyear{2001}}%
}]{%
dayan-abbott:01}%
\APACinsertmetastar{%
dayan-abbott:01}%
Dayan, P.%
\BCBT{}\ \BBA{} Abbott, L\BPBI F.%
%
\unskip\
\newblock
\APACrefYear{2001}.
\newblock
\APACrefbtitle{Theoretical Neuroscience : Computational and Mathematical
  Modeling of Neural Systems}{Theoretical neuroscience : Computational and
  mathematical modeling of neural systems}.
\newblock
\APACaddressPublisher{}{MIT Press}.
\PrintBackRefs{\CurrentBib}

\bibitem[\protect\citeauthoryear{%
Delorme%
, Perrinet%
\BCBL{}\ \BBA{} Thorpe%
}{%
Delorme%
\ \protect\BOthers{.}}{%
{\protect\APACyear{2001}}%
}]{%
delorme-perrinet-etal:01}%
\APACinsertmetastar{%
delorme-perrinet-etal:01}%
Delorme, A.%
, Perrinet, L.%
\BCBL{}\ \BBA{} Thorpe, S.%
%
\unskip\
\newblock
\APACrefYearMonthDay{2001}{}{}.
\newblock
\BBOQ{}\APACrefatitle{Network of integrate-and-fire neurons using Rank Order
  Coding B: spike timing dependant plasticity and emergence of orientation
  selectivity}{Network of integrate-and-fire neurons using rank order coding b:
  spike timing dependant plasticity and emergence of orientation
  selectivity}.\BBCQ{}
\newblock
\APACjournalVolNumPages{Neurocomputing}{38}{}{539--545}.
\PrintBackRefs{\CurrentBib}

\bibitem[\protect\citeauthoryear{%
Destexhe%
}{%
Destexhe%
}{%
{\protect\APACyear{1997}}%
}]{%
destexhe:97}%
\APACinsertmetastar{%
destexhe:97}%
Destexhe, A.%
%
\unskip\
\newblock
\APACrefYearMonthDay{1997}{}{}.
\newblock
\BBOQ{}\APACrefatitle{Conductance-Based Integrate and Fire
  Models}{Conductance-based integrate and fire models}.\BBCQ{}
\newblock
\APACjournalVolNumPages{Neural Computation}{9}{}{503--514}.
\PrintBackRefs{\CurrentBib}

\bibitem[\protect\citeauthoryear{%
Destexhe%
, Rudolph%
\BCBL{}\ \BBA{} Par\'e%
}{%
Destexhe%
\ \protect\BOthers{.}}{%
{\protect\APACyear{2003}}%
}]{%
destexhe-rudolph-etal:03}%
\APACinsertmetastar{%
destexhe-rudolph-etal:03}%
Destexhe, A.%
, Rudolph, M.%
\BCBL{}\ \BBA{} Par\'e, D.%
%
\unskip\
\newblock
\APACrefYearMonthDay{2003}{}{}.
\newblock
\BBOQ{}\APACrefatitle{The high-conductance state of neocortical neurons in
  vivo}{The high-conductance state of neocortical neurons in vivo}.\BBCQ{}
\newblock
\APACjournalVolNumPages{Nature Reviews Neuroscience}{4}{}{739--751}.
\PrintBackRefs{\CurrentBib}

\bibitem[\protect\citeauthoryear{%
Fr\'egnac%
}{%
Fr\'egnac%
}{%
{\protect\APACyear{2003}}%
}]{%
fregnac:03}%
\APACinsertmetastar{%
fregnac:03}%
Fr\'egnac, Y.%
%
\unskip\
\newblock
\APACrefYearMonthDay{2003}{}{}.
\newblock
\BBOQ{}\APACrefatitle{Association field in visual cortical neurons: From
  subthreshold visual synaptic integration to apparent-motion
  perception}{Association field in visual cortical neurons: From subthreshold
  visual synaptic integration to apparent-motion perception}.\BBCQ{}
\newblock
\BIn{} \APACrefbtitle{{E}uropean {C}onference on {V}isual {P}erception,
  {P}aris.}{{E}uropean {C}onference on {V}isual {P}erception, {P}aris.}
\PrintBackRefs{\CurrentBib}

\bibitem[\protect\citeauthoryear{%
Fr\'egnac%
}{%
Fr\'egnac%
}{%
{\protect\APACyear{2004}}%
}]{%
fregnac:04}%
\APACinsertmetastar{%
fregnac:04}%
Fr\'egnac, Y.%
%
\unskip\
\newblock
\APACrefYearMonthDay{2004}{}{}.
\newblock
\BBOQ{}\APACrefatitle{From synaptic rumours to low-level perception: an
  intracellular view of visual cortical dynamics}{From synaptic rumours to
  low-level perception: an intracellular view of visual cortical
  dynamics}.\BBCQ{}
\newblock
\APACjournalVolNumPages{Progress in Biochemistry and Biophysics}{31}{}{6--8}.
\PrintBackRefs{\CurrentBib}

\bibitem[\protect\citeauthoryear{%
Galarreta%
\ \BBA{} Hestrin%
}{%
Galarreta%
\ \BBA{} Hestrin%
}{%
{\protect\APACyear{2001}}%
}]{%
galarreta-hestin:01}%
\APACinsertmetastar{%
galarreta-hestin:01}%
Galarreta, M.%
\BCBT{}\ \BBA{} Hestrin, S.%
%
\unskip\
\newblock
\APACrefYearMonthDay{2001}{}{}.
\newblock
\BBOQ{}\APACrefatitle{Electrical Synapses Between GABA-Releasing
  Interneurons}{Electrical synapses between gaba-releasing
  interneurons}.\BBCQ{}
\newblock
\APACjournalVolNumPages{Nature Reviews Neuroscience}{2}{}{425--433}.
\PrintBackRefs{\CurrentBib}

\bibitem[\protect\citeauthoryear{%
Gautrais%
\ \BBA{} Thorpe%
}{%
Gautrais%
\ \BBA{} Thorpe%
}{%
{\protect\APACyear{1998}}%
}]{%
gautrais-thorpe:98}%
\APACinsertmetastar{%
gautrais-thorpe:98}%
Gautrais, J.%
\BCBT{}\ \BBA{} Thorpe, S.%
%
\unskip\
\newblock
\APACrefYearMonthDay{1998}{}{}.
\newblock
\BBOQ{}\APACrefatitle{Rate Coding vs Temporal Order Coding : a theorical
  approach}{Rate coding vs temporal order coding : a theorical
  approach}.\BBCQ{}
\newblock
\APACjournalVolNumPages{Biosystems}{48}{}{57--65}.
\PrintBackRefs{\CurrentBib}

\bibitem[\protect\citeauthoryear{%
Gerstein%
\ \BBA{} Mandelbrot%
}{%
Gerstein%
\ \BBA{} Mandelbrot%
}{%
{\protect\APACyear{1964}}%
}]{%
gerstein-mandelbrot:64}%
\APACinsertmetastar{%
gerstein-mandelbrot:64}%
Gerstein, G\BPBI L.%
\BCBT{}\ \BBA{} Mandelbrot, B.%
%
\unskip\
\newblock
\APACrefYearMonthDay{1964}{}{}.
\newblock
\BBOQ{}\APACrefatitle{Random walk models for the spike activity of a single
  neuron}{Random walk models for the spike activity of a single neuron}.\BBCQ{}
\newblock
\APACjournalVolNumPages{Biophysical Journal}{4}{}{41--68}.
\PrintBackRefs{\CurrentBib}

\bibitem[\protect\citeauthoryear{%
Gerstner%
\ \BBA{} Kistler%
}{%
Gerstner%
\ \BBA{} Kistler%
}{%
{\protect\APACyear{2002}}%
{\protect\APACexlab{{\protect\BCnt{2}}}}}]{%
gerstner-kistler:02b}%
\APACinsertmetastar{%
gerstner-kistler:02b}%
Gerstner, W.%
\BCBT{}\ \BBA{} Kistler, W.%
%
\unskip\
\newblock
\APACrefYear{2002{\protect\BCnt{2}}}.
\newblock
\APACrefbtitle{Spiking Neuron Models}{Spiking neuron models}.
\newblock
\APACaddressPublisher{}{Cambridge University Press}.
\PrintBackRefs{\CurrentBib}

\bibitem[\protect\citeauthoryear{%
Gerstner%
\ \BBA{} Kistler%
}{%
Gerstner%
\ \BBA{} Kistler%
}{%
{\protect\APACyear{2002}}%
{\protect\APACexlab{{\protect\BCnt{1}}}}}]{%
gerstner-kistler:02}%
\APACinsertmetastar{%
gerstner-kistler:02}%
Gerstner, W.%
\BCBT{}\ \BBA{} Kistler, W\BPBI M.%
%
\unskip\
\newblock
\APACrefYearMonthDay{2002{\protect\BCnt{1}}}{}{}.
\newblock
\BBOQ{}\APACrefatitle{Mathematical formulations of Hebbian
  learning.}{Mathematical formulations of hebbian learning.}\BBCQ{}
\newblock
\APACjournalVolNumPages{Biological Cybernetics}{87}{}{404--415}.
\PrintBackRefs{\CurrentBib}

\bibitem[\protect\citeauthoryear{%
Guyonneau%
, vanRullen%
\BCBL{}\ \BBA{} Thorpe%
}{%
Guyonneau%
\ \protect\BOthers{.}}{%
{\protect\APACyear{2004}}%
}]{%
guyonneau-vanrullen-etal:04}%
\APACinsertmetastar{%
guyonneau-vanrullen-etal:04}%
Guyonneau, R.%
, vanRullen, R.%
\BCBL{}\ \BBA{} Thorpe, S.%
%
\unskip\
\newblock
\APACrefYearMonthDay{2004}{}{}.
\newblock
\BBOQ{}\APACrefatitle{Neurons tune to the earliest spikes through
  STDP.}{Neurons tune to the earliest spikes through stdp.}\BBCQ{}
\newblock
\APACjournalVolNumPages{Neural Computation}{}{}{}.
\newblock
\APACrefnote{In review}
\PrintBackRefs{\CurrentBib}

\bibitem[\protect\citeauthoryear{%
Hodgkin%
\ \BBA{} Huxley%
}{%
Hodgkin%
\ \BBA{} Huxley%
}{%
{\protect\APACyear{1952}}%
}]{%
hodgkin-huxley:52}%
\APACinsertmetastar{%
hodgkin-huxley:52}%
Hodgkin, A.%
\BCBT{}\ \BBA{} Huxley, A.%
%
\unskip\
\newblock
\APACrefYearMonthDay{1952}{}{}.
\newblock
\BBOQ{}\APACrefatitle{A quantitative description of membrane current and its
  application to conduction and excitation in nerve.}{A quantitative
  description of membrane current and its application to conduction and
  excitation in nerve.}\BBCQ{}
\newblock
\APACjournalVolNumPages{Journal of Physiology}{117}{}{500--544}.
\PrintBackRefs{\CurrentBib}

\bibitem[\protect\citeauthoryear{%
Hornik%
, Stinchcombe%
\BCBL{}\ \BBA{} White%
}{%
Hornik%
\ \protect\BOthers{.}}{%
{\protect\APACyear{1989}}%
}]{%
hornik-etal:89}%
\APACinsertmetastar{%
hornik-etal:89}%
Hornik, K.%
, Stinchcombe, M.%
\BCBL{}\ \BBA{} White, H.%
%
\unskip\
\newblock
\APACrefYearMonthDay{1989}{}{}.
\newblock
\BBOQ{}\APACrefatitle{Multilayer feedforward networks are universal
  approximators}{Multilayer feedforward networks are universal
  approximators}.\BBCQ{}
\newblock
\APACjournalVolNumPages{Neural Networks}{2}{}{359--366}.
\PrintBackRefs{\CurrentBib}

\bibitem[\protect\citeauthoryear{%
E.~Izhikevich%
}{%
E.~Izhikevich%
}{%
{\protect\APACyear{2003}}%
}]{%
izhikevich:03}%
\APACinsertmetastar{%
izhikevich:03}%
Izhikevich, E.%
%
\unskip\
\newblock
\APACrefYearMonthDay{2003}{}{}.
\newblock
\BBOQ{}\APACrefatitle{Simple Model of Spiking Neurons}{Simple model of spiking
  neurons}.\BBCQ{}
\newblock
\APACjournalVolNumPages{IEEE Transactions on Neural
  Networks}{14}{6}{1569--1572}.
\PrintBackRefs{\CurrentBib}

\bibitem[\protect\citeauthoryear{%
E.~Izhikevich%
}{%
E.~Izhikevich%
}{%
{\protect\APACyear{2004}}%
}]{%
izhikevich:04}%
\APACinsertmetastar{%
izhikevich:04}%
Izhikevich, E.%
%
\unskip\
\newblock
\APACrefYearMonthDay{2004}{September}{}.
\newblock
\BBOQ{}\APACrefatitle{Which model to use for cortical spiking neurons?}{Which
  model to use for cortical spiking neurons?}\BBCQ{}
\newblock
\APACjournalVolNumPages{IEEE Trans Neural Netw}{15}{5}{1063--1070}.
\PrintBackRefs{\CurrentBib}

\bibitem[\protect\citeauthoryear{%
E\BPBI M.~Izhikevich%
}{%
E\BPBI M.~Izhikevich%
}{%
{\protect\APACyear{2006}}%
}]{%
izhikevich:06}%
\APACinsertmetastar{%
izhikevich:06}%
Izhikevich, E\BPBI M.%
%
\unskip\
\newblock
\APACrefYear{2006}.
\newblock
\APACrefbtitle{Dynamical Systems in Neuroscience: The Geometry of Excitability
  and Bursting}{Dynamical systems in neuroscience: The geometry of excitability
  and bursting}.
\newblock
\APACaddressPublisher{}{The MIT Press}.
\newblock
\APACrefnote{To appear.}
\PrintBackRefs{\CurrentBib}

\bibitem[\protect\citeauthoryear{%
Jaeger%
}{%
Jaeger%
}{%
{\protect\APACyear{2003}}%
}]{%
jaeger:03}%
\APACinsertmetastar{%
jaeger:03}%
Jaeger, H.%
%
\unskip\
\newblock
\APACrefYearMonthDay{2003}{}{}.
\newblock
\BBOQ{}\APACrefatitle{Adaptive nonlinear system identification with {E}cho
  {S}tate {N}etworks}{Adaptive nonlinear system identification with {E}cho
  {S}tate {N}etworks}.\BBCQ{}
\newblock
\BIn{} S.~Becker, S.~Thrun\BCBL{}\ \BBA{} K.~Obermayer\ (\BEDS),
  \APACrefbtitle{NIPS*2002, Advances in Neural Information Processing
  Systems}{Nips*2002, advances in neural information processing systems}\
  (\BVOL~15, \BPGS\ 593--600).
\newblock
\APACaddressPublisher{}{MIT Press}.
\PrintBackRefs{\CurrentBib}

\bibitem[\protect\citeauthoryear{%
Katok%
\ \BBA{} Hasselblatt%
}{%
Katok%
\ \BBA{} Hasselblatt%
}{%
{\protect\APACyear{1998}}%
}]{%
katok-hasselblatt:98}%
\APACinsertmetastar{%
katok-hasselblatt:98}%
Katok, A.%
\BCBT{}\ \BBA{} Hasselblatt, B.%
%
\unskip\
\newblock
\APACrefYear{1998}.
\newblock
\APACrefbtitle{Introduction to the modern theory of dynamical
  systems}{Introduction to the modern theory of dynamical systems}.
\newblock
\APACaddressPublisher{}{Kluwer}.
\PrintBackRefs{\CurrentBib}

\bibitem[\protect\citeauthoryear{%
Koch%
}{%
Koch%
}{%
{\protect\APACyear{1999}}%
}]{%
koch:99b}%
\APACinsertmetastar{%
koch:99b}%
Koch, C.%
%
\unskip\
\newblock
\APACrefYear{1999}.
\newblock
\APACrefbtitle{Biophysics of Computation: Information Processing in Single
  Neurons}{Biophysics of computation: Information processing in single
  neurons}.
\newblock
\APACaddressPublisher{}{Oxford University Press: New York.}
\PrintBackRefs{\CurrentBib}

\bibitem[\protect\citeauthoryear{%
Koch%
\ \BBA{} Segev%
}{%
Koch%
\ \BBA{} Segev%
}{%
{\protect\APACyear{1998}}%
}]{%
koch-segev:98}%
\APACinsertmetastar{%
koch-segev:98}%
Koch, C.%
\BCBT{}\ \BBA{} Segev, I.%
\ (\BEDS).
\unskip\
\newblock
\APACrefYear{1998}.
\newblock
\APACrefbtitle{Methods in Neuronal Modeling: From Ions to Networks}{Methods in
  neuronal modeling: From ions to networks}.
\newblock
\APACaddressPublisher{}{The MIT Press}.
\PrintBackRefs{\CurrentBib}

\bibitem[\protect\citeauthoryear{%
Kreuz%
, Haas%
, Morelli%
, Abarbanel%
\BCBL{}\ \BBA{} Politi%
}{%
Kreuz%
\ \protect\BOthers{.}}{%
{\protect\APACyear{2007}}%
}]{%
kreuz-haas:07}%
\APACinsertmetastar{%
kreuz-haas:07}%
Kreuz, T.%
, Haas, J\BPBI S.%
, Morelli, A.%
, Abarbanel, H\BPBI D.%
\BCBL{}\ \BBA{} Politi, A.%
%
\unskip\
\newblock
\APACrefYearMonthDay{2007}{}{}.
\newblock
\BBOQ{}\APACrefatitle{Measuring spike train synchrony and
  reliability}{Measuring spike train synchrony and reliability}.\BBCQ{}
\newblock
\BIn{} \APACrefbtitle{Computational Neurosciences meeting (CNS).}{Computational
  neurosciences meeting (cns).}
\PrintBackRefs{\CurrentBib}

\bibitem[\protect\citeauthoryear{%
Lazar%
}{%
Lazar%
}{%
{\protect\APACyear{2005}}%
}]{%
lazar:05}%
\APACinsertmetastar{%
lazar:05}%
Lazar, A.%
%
\unskip\
\newblock
\APACrefYearMonthDay{2005}{}{}.
\newblock
\BBOQ{}\APACrefatitle{Multichannel time encoding with integrate-and-fire
  neurons}{Multichannel time encoding with integrate-and-fire neurons}.\BBCQ{}
\newblock
\APACjournalVolNumPages{Neurocomputing}{65}{}{401--407}.
\PrintBackRefs{\CurrentBib}

\bibitem[\protect\citeauthoryear{%
Lewis%
\ \BBA{} Rinzel%
}{%
Lewis%
\ \BBA{} Rinzel%
}{%
{\protect\APACyear{2003}}%
}]{%
lewis-rinzel:03}%
\APACinsertmetastar{%
lewis-rinzel:03}%
Lewis, T\BPBI J.%
\BCBT{}\ \BBA{} Rinzel, J.%
%
\unskip\
\newblock
\APACrefYearMonthDay{2003}{}{}.
\newblock
\BBOQ{}\APACrefatitle{Dynamics of Spiking Neurons Connected by Both Inhibitory
  and Electrical Coupling.}{Dynamics of spiking neurons connected by both
  inhibitory and electrical coupling.}\BBCQ{}
\newblock
\APACjournalVolNumPages{Journal of Computational Neuroscience}{14}{3}{283-309}.
\PrintBackRefs{\CurrentBib}

\bibitem[\protect\citeauthoryear{%
Maass%
}{%
Maass%
}{%
{\protect\APACyear{1997}}%
}]{%
maass:97}%
\APACinsertmetastar{%
maass:97}%
Maass, W.%
%
\unskip\
\newblock
\APACrefYearMonthDay{1997}{}{}.
\newblock
\BBOQ{}\APACrefatitle{Fast sigmoidal networks via spiking neurons.}{Fast
  sigmoidal networks via spiking neurons.}\BBCQ{}
\newblock
\APACjournalVolNumPages{Neural Computation}{9}{}{279--304}.
\PrintBackRefs{\CurrentBib}

\bibitem[\protect\citeauthoryear{%
Maass%
}{%
Maass%
}{%
{\protect\APACyear{2001}}%
}]{%
maass:01}%
\APACinsertmetastar{%
maass:01}%
Maass, W.%
%
\unskip\
\newblock
\APACrefYearMonthDay{2001}{}{}.
\newblock
\BBOQ{}\APACrefatitle{On the relevance of time in neural computation and
  learning}{On the relevance of time in neural computation and
  learning}.\BBCQ{}
\newblock
\APACjournalVolNumPages{Theoretical Computer Science}{261}{}{157--178}.
\newblock
\APACrefnote{(extended version of ALT'97, in LNAI 1316:364-384)}
\PrintBackRefs{\CurrentBib}

\bibitem[\protect\citeauthoryear{%
Maass%
\ \BBA{} Bishop%
}{%
Maass%
\ \BBA{} Bishop%
}{%
{\protect\APACyear{2003}}%
}]{%
maass-bishop:03}%
\APACinsertmetastar{%
maass-bishop:03}%
Maass, W.%
\BCBT{}\ \BBA{} Bishop, C\BPBI M.%
\ (\BEDS).
\unskip\
\newblock
\APACrefYear{2003}.
\newblock
\APACrefbtitle{Pulsed Neural Networks}{Pulsed neural networks}.
\newblock
\APACaddressPublisher{}{MIT Press}.
\PrintBackRefs{\CurrentBib}

\bibitem[\protect\citeauthoryear{%
Maass%
\ \BBA{} Natschlager%
}{%
Maass%
\ \BBA{} Natschlager%
}{%
{\protect\APACyear{1997}}%
}]{%
maass-natschlager:97}%
\APACinsertmetastar{%
maass-natschlager:97}%
Maass, W.%
\BCBT{}\ \BBA{} Natschlager, T.%
%
\unskip\
\newblock
\APACrefYearMonthDay{1997}{}{}.
\newblock
\BBOQ{}\APACrefatitle{Networks of Spiking Neurons can Emulate Arbitrary
  Hopfield nets in Temporal Coding}{Networks of spiking neurons can emulate
  arbitrary hopfield nets in temporal coding}.\BBCQ{}
\newblock
\APACjournalVolNumPages{Neural Systems}{8}{4}{355--372}.
\PrintBackRefs{\CurrentBib}

\bibitem[\protect\citeauthoryear{%
Maass%
, Natschl\"ager%
\BCBL{}\ \BBA{} Markram%
}{%
Maass%
\ \protect\BOthers{.}}{%
{\protect\APACyear{2002}}%
}]{%
maass-etal:02}%
\APACinsertmetastar{%
maass-etal:02}%
Maass, W.%
, Natschl\"ager, T.%
\BCBL{}\ \BBA{} Markram, H.%
%
\unskip\
\newblock
\APACrefYearMonthDay{2002}{}{}.
\newblock
\BBOQ{}\APACrefatitle{Real-time computing without stable states: A new
  framework for neural computation based on perturbations}{Real-time computing
  without stable states: A new framework for neural computation based on
  perturbations}.\BBCQ{}
\newblock
\APACjournalVolNumPages{Neural Computation}{14}{11}{2531--2560}.
\PrintBackRefs{\CurrentBib}

\bibitem[\protect\citeauthoryear{%
Mainen%
\ \BBA{} Sejnowski%
}{%
Mainen%
\ \BBA{} Sejnowski%
}{%
{\protect\APACyear{1995}}%
}]{%
mainen-sejnowski:95}%
\APACinsertmetastar{%
mainen-sejnowski:95}%
Mainen, Z.%
\BCBT{}\ \BBA{} Sejnowski, T.%
%
\unskip\
\newblock
\APACrefYearMonthDay{1995}{}{}.
\newblock
\BBOQ{}\APACrefatitle{Reliability of spike timing in neocortical
  neurons}{Reliability of spike timing in neocortical neurons}.\BBCQ{}
\newblock
\APACjournalVolNumPages{Science}{268}{5216}{1503-1506}.
\PrintBackRefs{\CurrentBib}

\bibitem[\protect\citeauthoryear{%
Markram%
, L\"{u}bke%
, Frotscher%
\BCBL{}\ \BBA{} Sakmann%
}{%
Markram%
\ \protect\BOthers{.}}{%
{\protect\APACyear{1997}}%
}]{%
markram-etal:97}%
\APACinsertmetastar{%
markram-etal:97}%
Markram, H.%
, L\"{u}bke, J.%
, Frotscher, M.%
\BCBL{}\ \BBA{} Sakmann, B.%
%
\unskip\
\newblock
\APACrefYearMonthDay{1997}{}{}.
\newblock
\BBOQ{}\APACrefatitle{Regulation of synaptic efficacy by coincidence of
  postsynaptic AP and EPSP}{Regulation of synaptic efficacy by coincidence of
  postsynaptic ap and epsp}.\BBCQ{}
\newblock
\APACjournalVolNumPages{Science}{275}{213}{}.
\PrintBackRefs{\CurrentBib}

\bibitem[\protect\citeauthoryear{%
Martinez%
\ \BBA{} Paugam-Moisy%
}{%
Martinez%
\ \BBA{} Paugam-Moisy%
}{%
{\protect\APACyear{2008}}%
}]{%
martinez-paugam:08}%
\APACinsertmetastar{%
martinez-paugam:08}%
Martinez, R.%
\BCBT{}\ \BBA{} Paugam-Moisy, H.%
%
\unskip\
\newblock
\APACrefYearMonthDay{2008}{}{}.
\newblock
\BBOQ{}\APACrefatitle{Les groupes polychrones pour capturer l'aspect
  spatio-temporel de la m\'emorisation}{Les groupes polychrones pour capturer
  l'aspect spatio-temporel de la m\'emorisation}.\BBCQ{}
\newblock
\BIn{} \APACrefbtitle{Neurocomp 2008.}{Neurocomp 2008.}
\PrintBackRefs{\CurrentBib}

\bibitem[\protect\citeauthoryear{%
McCormick%
\ \BBA{} Bal%
}{%
McCormick%
\ \BBA{} Bal%
}{%
{\protect\APACyear{1997}}%
}]{%
macormick-bal:97}%
\APACinsertmetastar{%
macormick-bal:97}%
McCormick, D\BPBI A.%
\BCBT{}\ \BBA{} Bal, T.%
%
\unskip\
\newblock
\APACrefYearMonthDay{1997}{}{}.
\newblock
\BBOQ{}\APACrefatitle{Sleep and Arousal: Thalamocortical Mechanisms}{Sleep and
  arousal: Thalamocortical mechanisms}.\BBCQ{}
\newblock
\APACjournalVolNumPages{Annual Review of Neuroscience}{20}{}{185-215}.
\PrintBackRefs{\CurrentBib}

\bibitem[\protect\citeauthoryear{%
Morrison%
, Mehring%
, Geisel%
, Aerstsen%
\BCBL{}\ \BBA{} Diesmann%
}{%
Morrison%
\ \protect\BOthers{.}}{%
{\protect\APACyear{2005}}%
}]{%
morrison-mehring-etal:05}%
\APACinsertmetastar{%
morrison-mehring-etal:05}%
Morrison, A.%
, Mehring, C.%
, Geisel, T.%
, Aerstsen, A.%
\BCBL{}\ \BBA{} Diesmann, M.%
%
\unskip\
\newblock
\APACrefYearMonthDay{2005}{}{}.
\newblock
\BBOQ{}\APACrefatitle{Advancing the boundaries of high connectivity network
  with distributed computing}{Advancing the boundaries of high connectivity
  network with distributed computing}.\BBCQ{}
\newblock
\APACjournalVolNumPages{Neural Comput}{17}{8}{1776--1801}.
\PrintBackRefs{\CurrentBib}

\bibitem[\protect\citeauthoryear{%
Paré%
, Bouhassira%
, Oakson%
\BCBL{}\ \BBA{} Datta%
}{%
Paré%
\ \protect\BOthers{.}}{%
{\protect\APACyear{1990}}%
}]{%
pare-bouhassira-etal:90}%
\APACinsertmetastar{%
pare-bouhassira-etal:90}%
Paré, D.%
, Bouhassira, D.%
, Oakson, G.%
\BCBL{}\ \BBA{} Datta, S.%
%
\unskip\
\newblock
\APACrefYearMonthDay{1990}{}{}.
\newblock
\BBOQ{}\APACrefatitle{Spontaneous and evoked activities of anterior thalamic
  neurons during waking and sleep states}{Spontaneous and evoked activities of
  anterior thalamic neurons during waking and sleep states}.\BBCQ{}
\newblock
\APACjournalVolNumPages{Experimental Brain Research}{80}{1}{}.
\PrintBackRefs{\CurrentBib}

\bibitem[\protect\citeauthoryear{%
Paugam-Moisy%
\ \BBA{} Bohte%
}{%
Paugam-Moisy%
\ \BBA{} Bohte%
}{%
{\protect\APACyear{2009}}%
}]{%
paugam-bohte:09}%
\APACinsertmetastar{%
paugam-bohte:09}%
Paugam-Moisy, H.%
\BCBT{}\ \BBA{} Bohte, S.%
%
\unskip\
\newblock
\APACrefYearMonthDay{2009}{}{}.
\newblock
\BBOQ{}\APACrefatitle{Handbook of Natural Computing}{Handbook of natural
  computing}.\BBCQ{}
\newblock
\BIn{} J.~Kok\ \BBA{} T.~Heskes\ (\BEDS), (\BCHAP\ Computing with Spiking
  Neuron Networks).
\newblock
\APACaddressPublisher{}{Springer Verlag}.
\newblock
\APACrefnote{(to appear)}
\PrintBackRefs{\CurrentBib}

\bibitem[\protect\citeauthoryear{%
Paugam-Moisy%
, Martinez%
\BCBL{}\ \BBA{} Bengio%
}{%
Paugam-Moisy%
\ \protect\BOthers{.}}{%
{\protect\APACyear{2008}}%
}]{%
paugam-moisy-etal:08}%
\APACinsertmetastar{%
paugam-moisy-etal:08}%
Paugam-Moisy, H.%
, Martinez, R.%
\BCBL{}\ \BBA{} Bengio, S.%
%
\unskip\
\newblock
\APACrefYearMonthDay{2008}{}{}.
\newblock
\BBOQ{}\APACrefatitle{Delay learning and polychronization for reservoir
  computing}{Delay learning and polychronization for reservoir
  computing}.\BBCQ{}
\newblock
\BIn{} (\BVOL~71, \BPGS\ 1143--1158).
\PrintBackRefs{\CurrentBib}

\bibitem[\protect\citeauthoryear{%
Pfister%
\ \BBA{} Gerstner%
}{%
Pfister%
\ \BBA{} Gerstner%
}{%
{\protect\APACyear{2006}}%
}]{%
pfister-gerstner:06}%
\APACinsertmetastar{%
pfister-gerstner:06}%
Pfister, J\BHBI P.%
\BCBT{}\ \BBA{} Gerstner, W.%
%
\unskip\
\newblock
\APACrefYearMonthDay{2006}{}{}.
\newblock
\BBOQ{}\APACrefatitle{Triplets of Spikes in a Model of Spike Timing-Dependent
  Plasticity}{Triplets of spikes in a model of spike timing-dependent
  plasticity}.\BBCQ{}
\newblock
\APACjournalVolNumPages{J. Neurosci.}{26}{}{9673--9682}.
\newblock
 \begin{APACrefURL}
  \url{http://icwww.epfl.ch/~gerstner//PUBLICATIONS/Pfister06b.pdf}
  \end{APACrefURL}
\PrintBackRefs{\CurrentBib}

\bibitem[\protect\citeauthoryear{%
Politi%
\ \BBA{} Torcini%
}{%
Politi%
\ \BBA{} Torcini%
}{%
{\protect\APACyear{2009}}%
}]{%
politi-torcini:09}%
\APACinsertmetastar{%
politi-torcini:09}%
Politi, A.%
\BCBT{}\ \BBA{} Torcini, A.%
%
\unskip\
\newblock
\APACrefYearMonthDay{2009}{}{}.
\newblock
\BBOQ{}\APACrefatitle{Stable chaos}{Stable chaos}.\BBCQ{}
\newblock
\APACjournalVolNumPages{http://lanl.arxiv.org/abs/0902.2545}{}{}{}.
\PrintBackRefs{\CurrentBib}

\bibitem[\protect\citeauthoryear{%
Rauch%
, La~Camera%
, Luscher%
, Senn%
\BCBL{}\ \BBA{} Fusi%
}{%
Rauch%
\ \protect\BOthers{.}}{%
{\protect\APACyear{2003}}%
}]{%
rauch-etal:03}%
\APACinsertmetastar{%
rauch-etal:03}%
Rauch, A.%
, La~Camera, G.%
, Luscher, H\BHBI R.%
, Senn, W.%
\BCBL{}\ \BBA{} Fusi, S.%
%
\unskip\
\newblock
\APACrefYearMonthDay{2003}{}{}.
\newblock
\BBOQ{}\APACrefatitle{Neocortical Pyramidal Cells Respond as Integrate-and-Fire
  Neurons to in Vivo-Like Input Currents}{Neocortical pyramidal cells respond
  as integrate-and-fire neurons to in vivo-like input currents}.\BBCQ{}
\newblock
\APACjournalVolNumPages{J Neurophysiol}{90}{3}{1598-1612}.
\PrintBackRefs{\CurrentBib}

\bibitem[\protect\citeauthoryear{%
Rieke%
, Warland%
, Steveninck%
\BCBL{}\ \BBA{} Bialek%
}{%
Rieke%
\ \protect\BOthers{.}}{%
{\protect\APACyear{1996}}%
}]{%
rieke-etal:96}%
\APACinsertmetastar{%
rieke-etal:96}%
Rieke, F.%
, Warland, D.%
, Steveninck, R. de~Ruyter~van%
\BCBL{}\ \BBA{} Bialek, W.%
%
\unskip\
\newblock
\APACrefYear{1996}.
\newblock
\APACrefbtitle{Spikes, Exploring the Neural Code}{Spikes, exploring the neural
  code}.
\newblock
\APACaddressPublisher{}{The M.I.T. Press}.
\PrintBackRefs{\CurrentBib}

\bibitem[\protect\citeauthoryear{%
Rostro-Gonzalez%
, Cessac%
, Vasquez%
\BCBL{}\ \BBA{} Vi\'eville%
}{%
Rostro-Gonzalez%
\ \protect\BOthers{.}}{%
{\protect\APACyear{2009}}%
}]{%
rostro-etal:09}%
\APACinsertmetastar{%
rostro-etal:09}%
Rostro-Gonzalez, H.%
, Cessac, B.%
, Vasquez, J\BPBI C.%
\BCBL{}\ \BBA{} Vi\'eville, T.%
%
\unskip\
\newblock
\APACrefYearMonthDay{2009}{}{}.
\newblock
\BBOQ{}\APACrefatitle{Back-engineering of spiking neural networks
  parameters}{Back-engineering of spiking neural networks parameters}.\BBCQ{}
\newblock
\BIn{} \APACrefbtitle{Computational Neurosciences meeting (CNS).}{Computational
  neurosciences meeting (cns).}
\PrintBackRefs{\CurrentBib}

\bibitem[\protect\citeauthoryear{%
Rudolph%
\ \BBA{} Destexhe%
}{%
Rudolph%
\ \BBA{} Destexhe%
}{%
{\protect\APACyear{2007}}%
}]{%
rudolph-destexhe:07}%
\APACinsertmetastar{%
rudolph-destexhe:07}%
Rudolph, M.%
\BCBT{}\ \BBA{} Destexhe, A.%
%
\unskip\
\newblock
\APACrefYearMonthDay{2007}{}{}.
\newblock
\BBOQ{}\APACrefatitle{How much can we trust neural simulation strategies?}{How
  much can we trust neural simulation strategies?}\BBCQ{}
\newblock
\APACjournalVolNumPages{Neurocomputing}{}{}{}.
\newblock
\APACrefnote{To appear}
\PrintBackRefs{\CurrentBib}

\bibitem[\protect\citeauthoryear{%
Sch\"afer%
\ \BBA{} Zimmermann%
}{%
Sch\"afer%
\ \BBA{} Zimmermann%
}{%
{\protect\APACyear{2006}}%
}]{%
schafer-zimmermann:06}%
\APACinsertmetastar{%
schafer-zimmermann:06}%
Sch\"afer, A\BPBI M.%
\BCBT{}\ \BBA{} Zimmermann, H\BPBI G.%
%
\unskip\
\newblock
\APACrefYearMonthDay{2006}{}{}.
\newblock
\BBOQ{}\APACrefatitle{Recurrent Neural Networks Are Universal
  Approximators}{Recurrent neural networks are universal approximators}.\BBCQ{}
\newblock
\APACjournalVolNumPages{Lecture Notes in Computer Science}{4131}{}{632--640}.
\newblock
 \begin{APACrefURL}
  \url{http://www.springerlink.com/content/5635187408g7k2x3/fulltext.pdf}
  \end{APACrefURL}
\PrintBackRefs{\CurrentBib}

\bibitem[\protect\citeauthoryear{%
Schrauwen%
}{%
Schrauwen%
}{%
{\protect\APACyear{2007}}%
}]{%
schrauwen:07}%
\APACinsertmetastar{%
schrauwen:07}%
Schrauwen, B.%
%
\unskip\
\newblock
\APACrefYear{2007}.
\newblock
\APACrefbtitle{Towards Applicable Spiking Neural Networks}{Towards applicable
  spiking neural networks}.
\newblock
\BUPhD, Universiteit Gent, Belgium.
\PrintBackRefs{\CurrentBib}

\bibitem[\protect\citeauthoryear{%
Shadlen%
\ \BBA{} Newsome%
}{%
Shadlen%
\ \BBA{} Newsome%
}{%
{\protect\APACyear{1994}}%
}]{%
shadlen-newsome:94}%
\APACinsertmetastar{%
shadlen-newsome:94}%
Shadlen, M\BPBI N.%
\BCBT{}\ \BBA{} Newsome, W\BPBI T.%
%
\unskip\
\newblock
\APACrefYearMonthDay{1994}{August}{}.
\newblock
\BBOQ{}\APACrefatitle{Noise, neural codes and cortical organization.}{Noise,
  neural codes and cortical organization.}\BBCQ{}
\newblock
\APACjournalVolNumPages{Curr Opin Neurobiol}{4}{4}{569--579}.
\newblock
 \begin{APACrefURL} \url{{http://www.ncbi.nlm.nih.gov/entrez/query.fcgi?cmd},
  owner = { jtouboul }} \end{APACrefURL}
\PrintBackRefs{\CurrentBib}

\bibitem[\protect\citeauthoryear{%
Simoncelli%
\ \BBA{} Olshausen%
}{%
Simoncelli%
\ \BBA{} Olshausen%
}{%
{\protect\APACyear{2001}}%
}]{%
simoncelli-olshausen:01}%
\APACinsertmetastar{%
simoncelli-olshausen:01}%
Simoncelli, E.%
\BCBT{}\ \BBA{} Olshausen, B.%
%
\unskip\
\newblock
\APACrefYearMonthDay{2001}{}{}.
\newblock
\BBOQ{}\APACrefatitle{Natural IMAGE STATISTICS AND NEURAL
  REPRESENTATION}{Natural image statistics and neural representation}.\BBCQ{}
\newblock
\APACjournalVolNumPages{Annual Review of Neuroscience}{24}{1}{1193--1216}.
\PrintBackRefs{\CurrentBib}

\bibitem[\protect\citeauthoryear{%
Thorpe%
, Delorme%
\BCBL{}\ \BBA{} VanRullen%
}{%
Thorpe%
\ \protect\BOthers{.}}{%
{\protect\APACyear{2001}}%
}]{%
thorpe-delorme-etal:01}%
\APACinsertmetastar{%
thorpe-delorme-etal:01}%
Thorpe, S.%
, Delorme, A.%
\BCBL{}\ \BBA{} VanRullen, R.%
%
\unskip\
\newblock
\APACrefYearMonthDay{2001}{}{}.
\newblock
\BBOQ{}\APACrefatitle{Spike based strategies for rapid processing.}{Spike based
  strategies for rapid processing.}\BBCQ{}
\newblock
\APACjournalVolNumPages{Neural Networks}{14}{}{715--726}.
\PrintBackRefs{\CurrentBib}

\bibitem[\protect\citeauthoryear{%
Thorpe%
\ \BBA{} Fabre-Thorpe%
}{%
Thorpe%
\ \BBA{} Fabre-Thorpe%
}{%
{\protect\APACyear{2001}}%
}]{%
thorpe-fabre-thorpe:01}%
\APACinsertmetastar{%
thorpe-fabre-thorpe:01}%
Thorpe, S.%
\BCBT{}\ \BBA{} Fabre-Thorpe, M.%
%
\unskip\
\newblock
\APACrefYearMonthDay{2001}{}{}.
\newblock
\BBOQ{}\APACrefatitle{Seeking categories in the brain}{Seeking categories in
  the brain}.\BBCQ{}
\newblock
\APACjournalVolNumPages{Science}{291}{}{260--263}.
\PrintBackRefs{\CurrentBib}

\bibitem[\protect\citeauthoryear{%
Touboul%
\ \BBA{} Brette%
}{%
Touboul%
\ \BBA{} Brette%
}{%
{\protect\APACyear{2008}}%
}]{%
touboul-brette:08}%
\APACinsertmetastar{%
touboul-brette:08}%
Touboul, J.%
\BCBT{}\ \BBA{} Brette, R.%
%
\unskip\
\newblock
\APACrefYearMonthDay{2008}{nov}{}.
\newblock
\BBOQ{}\APACrefatitle{Dynamics and bifurcations of the adaptive exponential
  integrate-and-fire model}{Dynamics and bifurcations of the adaptive
  exponential integrate-and-fire model}.\BBCQ{}
\newblock
\APACjournalVolNumPages{Biological Cybernetics}{99}{4--5}{319--334}.
\newblock
 \begin{APACrefURL} \url{http://www.ncbi.nlm.nih.gov/pubmed/19011921}
  \end{APACrefURL}
\newblock
\APACrefnote{PMID: 19011921 DOI: 10.1007/s00422-008-0267-4}
\PrintBackRefs{\CurrentBib}

\bibitem[\protect\citeauthoryear{%
Toyoizumi%
, Pfister%
, Aihara%
\BCBL{}\ \BBA{} Gerstner%
}{%
Toyoizumi%
\ \protect\BOthers{.}}{%
{\protect\APACyear{2005}}%
}]{%
toyoizumi-etal:05}%
\APACinsertmetastar{%
toyoizumi-etal:05}%
Toyoizumi, T.%
, Pfister, J\BHBI P.%
, Aihara, K.%
\BCBL{}\ \BBA{} Gerstner, W.%
%
\unskip\
\newblock
\APACrefYearMonthDay{2005}{}{}.
\newblock
\BBOQ{}\APACrefatitle{Generalized Bienenstock-Cooper-Munro rule for spiking
  neurons that maximizes information transmission}{Generalized
  bienenstock-cooper-munro rule for spiking neurons that maximizes information
  transmission}.\BBCQ{}
\newblock
\APACjournalVolNumPages{Proceedings of the National Academy of
  Science}{102}{}{5239--5244}.
\PrintBackRefs{\CurrentBib}

\bibitem[\protect\citeauthoryear{%
Toyoizumi%
, Pfister%
, Aihara%
\BCBL{}\ \BBA{} Gerstner%
}{%
Toyoizumi%
\ \protect\BOthers{.}}{%
{\protect\APACyear{2007}}%
}]{%
toyoizumi-etal:07}%
\APACinsertmetastar{%
toyoizumi-etal:07}%
Toyoizumi, T.%
, Pfister, J\BHBI P.%
, Aihara, K.%
\BCBL{}\ \BBA{} Gerstner, W.%
%
\unskip\
\newblock
\APACrefYearMonthDay{2007}{}{}.
\newblock
\BBOQ{}\APACrefatitle{Optimality Model of Unsupervised Spike-Timing Dependent
  Plasticity: Synaptic Memory and Weight Distribution}{Optimality model of
  unsupervised spike-timing dependent plasticity: Synaptic memory and weight
  distribution}.\BBCQ{}
\newblock
\APACjournalVolNumPages{Neural Computation}{19}{}{639--671}.
\PrintBackRefs{\CurrentBib}

\bibitem[\protect\citeauthoryear{%
Verstraeten%
, Schrauwen%
, D’Haene%
\BCBL{}\ \BBA{} Stroobandt%
}{%
Verstraeten%
\ \protect\BOthers{.}}{%
{\protect\APACyear{2007}}%
}]{%
verstraeten-etal:07}%
\APACinsertmetastar{%
verstraeten-etal:07}%
Verstraeten, D.%
, Schrauwen, B.%
, D’Haene, M.%
\BCBL{}\ \BBA{} Stroobandt, D.%
%
\unskip\
\newblock
\APACrefYearMonthDay{2007}{}{}.
\newblock
\BBOQ{}\APACrefatitle{An experimental unification of reservoir computing
  methods}{An experimental unification of reservoir computing methods}.\BBCQ{}
\newblock
\APACjournalVolNumPages{Neural Networks}{20}{3}{391--403}.
\PrintBackRefs{\CurrentBib}

\bibitem[\protect\citeauthoryear{%
Victor%
}{%
Victor%
}{%
{\protect\APACyear{2005}}%
}]{%
victor:05}%
\APACinsertmetastar{%
victor:05}%
Victor, J.%
%
\unskip\
\newblock
\APACrefYearMonthDay{2005}{}{}.
\newblock
\BBOQ{}\APACrefatitle{Spike train metrics}{Spike train metrics}.\BBCQ{}
\newblock
\APACjournalVolNumPages{Current Opinion in Neurobiology}{15}{5}{585--592}.
\PrintBackRefs{\CurrentBib}

\bibitem[\protect\citeauthoryear{%
Victor%
\ \BBA{} Purpura%
}{%
Victor%
\ \BBA{} Purpura%
}{%
{\protect\APACyear{1996}}%
}]{%
victor-purpura:96}%
\APACinsertmetastar{%
victor-purpura:96}%
Victor, J.%
\BCBT{}\ \BBA{} Purpura, K.%
%
\unskip\
\newblock
\APACrefYearMonthDay{1996}{}{}.
\newblock
\BBOQ{}\APACrefatitle{Nature and precision of temporal coding in visual cortex:
  a metric-space analysis.}{Nature and precision of temporal coding in visual
  cortex: a metric-space analysis.}\BBCQ{}
\newblock
\APACjournalVolNumPages{J Neurophysiol}{76}{}{1310--1326}.
\PrintBackRefs{\CurrentBib}

\bibitem[\protect\citeauthoryear{%
Vi\'eville%
\ \BBA{} Crahay%
}{%
Vi\'eville%
\ \BBA{} Crahay%
}{%
{\protect\APACyear{2004}}%
}]{%
vieville-crahay:04}%
\APACinsertmetastar{%
vieville-crahay:04}%
Vi\'eville, T.%
\BCBT{}\ \BBA{} Crahay, S.%
%
\unskip\
\newblock
\APACrefYearMonthDay{2004}{}{}.
\newblock
\BBOQ{}\APACrefatitle{Using an Hebbian Learning Rule for Multi-Class SVM
  Classifiers}{Using an hebbian learning rule for multi-class svm
  classifiers}.\BBCQ{}
\newblock
\APACjournalVolNumPages{Journal of Computational
  Neuroscience}{17}{3}{271--287}.
\newblock
 \begin{APACrefURL}
  \url{http://journals.kluweronline.com/article.asp?PIPS=5384399}
  \end{APACrefURL}
\PrintBackRefs{\CurrentBib}

\bibitem[\protect\citeauthoryear{%
\v{S}\'ima%
\ \BBA{} Sgall%
}{%
\v{S}\'ima%
\ \BBA{} Sgall%
}{%
{\protect\APACyear{2005}}%
}]{%
sima-sgall:05}%
\APACinsertmetastar{%
sima-sgall:05}%
\v{S}\'ima, J.%
\BCBT{}\ \BBA{} Sgall, J.%
%
\unskip\
\newblock
\APACrefYearMonthDay{2005}{}{}.
\newblock
\BBOQ{}\APACrefatitle{On the Nonlearnability of a Single Spiking Neuron}{On the
  nonlearnability of a single spiking neuron}.\BBCQ{}
\newblock
\APACjournalVolNumPages{Neural Computation}{17}{12}{2635--2647}.
\PrintBackRefs{\CurrentBib}

\end{thebibliography}

{\small {\bf Acknowledgment:} Partially supported by the ANR MAPS \& the MACCAC ARC projects.}

\end{document}